\documentclass[12pt]{article}
\input psfig.sty

\def\ie{{\it i.e.}}
\def\eg{{\it e.g.}}

\def\etal{{\it et al.}}

\def\sub#1{_{\lower.25ex\hbox{$\scriptstyle#1$}}}
\def\to{\rightarrow}

\newskip\zatskip \zatskip=0pt plus0pt minus0pt
\def\matth{\mathsurround=0pt}
\def\lsim{\mathrel{\mathpalette\atversim<}}
\def\gsim{\mathrel{\mathpalette\atversim>}}
\def\atversim#1#2{\lower0.7ex\vbox{\baselineskip\zatskip\lineskip\zatskip
  \lineskiplimit 0pt\ialign{$\matth#1\hfil##\hfil$\crcr#2\crcr\sim\crcr}}}

\renewcommand{\thefootnote}{\fnsymbol{footnote}}

\begin{document} \begin{titlepage} 
\rightline{\vbox{\halign{&#\hfil\cr
&SLAC-PUB-9585\cr
&November 2002\cr}}}
\begin{center} 
\openup4.0\jot
 
{\Large\bf Constraining the Littlest Higgs}
\footnote{Work supported by the Department of 
Energy, Contract DE-AC03-76SF00515}
\medskip

\normalsize 
JoAnne L. Hewett, Frank J. Petriello, and Thomas G. Rizzo
\vskip .2cm
Stanford Linear Accelerator Center \\
Stanford University \\
Stanford CA 94309, USA\\
\vskip .2cm

\end{center} 
\openup2.0\jot
\begin{abstract} 
Little Higgs models offer a new way to address the hierarchy problem,
and give rise to a weakly-coupled Higgs sector.  These 
theories predict the existence of new states which are necessary to cancel the 
quadratic divergences of the Standard Model. The simplest version of 
these models, the Littlest Higgs, is based on an $SU(5)/SO(5)$ 
non-linear sigma model  and predicts that
four new gauge bosons, a weak isosinglet quark, $t'$,  with $Q=2/3$, 
as well as an isotriplet scalar field exist at the TeV scale.
We consider the contributions of these new states to precision electroweak 
observables, and examine their production at the Tevatron. We  
thoroughly explore the parameter space of this model and find that small 
regions are allowed by the precision data where the model parameters take 
on their natural values.  These regions are, however, excluded by the
Tevatron data.  Combined, the direct 
and indirect effects of these new states constrain the `decay constant' 
$f\gsim 3.5$ TeV and $m_{t'}\gsim 7 $ TeV. These bounds imply that 
significant fine-tuning be present in order for this model to 
resolve the hierarchy problem.
\end{abstract}

\renewcommand{\thefootnote}{\arabic{footnote}} \end{titlepage}


\section{Introduction}
  
The Standard Model (SM) of electroweak physics is a remarkable achievement.
Precision electroweak experiments have probed the SM at the level of
quantum effects and have confirmed every feature of the theory.  In
particular, the set of precision electroweak (EW) data
has tested the SM beyond one-loop level and no significant deviations
from SM predictions have been observed.  The symmetry breaking sector 
of the SM has been investigated through its virtual effects and 
precision measurements strongly prefer the existence of a weakly coupled 
Higgs boson by constraining its mass
to be $m_H< 193$ GeV at $95\%$ CL \cite{ewfit}.  This upper bound may
be relaxed if certain classes of 
new physics contributions beyond the SM are present at the TeV scale 
and compensate for the indirect effects of a heavier scalar 
sector \cite{PW}.  However, because of decoupling, models with an 
elementary Higgs scalar typically yield small contributions to the
electroweak radiative corrections.  Direct evidence
for the electroweak symmetry breaking dynamics has yet to be observed, and
searches for Higgs boson production 
have placed the compatible lower bound $m_H>114.4$ GeV.

The existence of a weakly coupled Higgs sector generates the
hierarchy problem, which is a long-standing puzzle in particle physics.
Few candidates exist for the mechanism which solves this problem.
Supersymmetry naturally stabilizes the hierarchy, as
the quadratically divergent loop contributions to the Higgs mass cancel
between the fermion and boson contributions, provided the supersymmetry
breaking scale is near a TeV \cite{HK}.  
Novel theories with extra dimensions exploit the geometry of the
higher dimensional spacetime to resolve the hierarchy \cite{xdim} and result 
in numerous phenomenological consequences at the TeV scale \cite{xdimrev}.
Supersymmetry has the additional feature
of remaining weakly coupled up to high scales, whereas
quantum gravity becomes strong at the TeV scale in these extra-dimensional 
models.  Experimental confirmation of either scenario has yet to occur.

Recently, another scenario has been developed which attacks the hierarchy
problem in a new way, while maintaining a weakly coupled scalar sector.
This scenario is known as the Little Higgs \cite{hckn,others} where
the Higgs is effectively a pseudo-Goldstone boson.  These theories are
realizations of earlier attempts \cite{georgi} to stablize a Higgs
which arises as a pseudo-Goldstone boson resulting from a 
spontaneously broken non-linear approximate symmetry.
This approximate global symmetry protects the Higgs vev relative to
the UV cut-off of the theory, $\Lambda$, which appears at a higher scale.
The recent progress was
attained in models of dimensional deconstruction \cite{decon}, where
the quadratically divergent corrections to the Higgs mass were 
shown to cancel with contributions from new fields in
non-trivial descriptions of theory space.  The phenomenology of these
models was examined in \cite{deconphen}.  The recent Little Higgs
models have the advantage of not requiring a non-trivial theory
space in order for the quadratic divergences to be eliminated.
The novel feature of these cancellations, is that the divergent 
contributions from a particular particle are cancelled by a new particle
of the {\it same} spin.  Hence, these theories predict the existence of
new $Q=+2/3$ quarks, gauge bosons, and scalars,
all with masses at the TeV scale, in order to remove the relevant 
divergences.  In addition, a small mass for the
Higgs boson, of order 100 GeV, is naturally obtained from multi-loop 
corrections.  

The most economical model of this type to 
date, known as the Littlest Higgs \cite{hckn},
is based on an $SU(5)/SO(5)$ nonlinear sigma model.  
The UV cut-off of this theory occurs at $\Lambda\sim 4\pi f\sim 10$ TeV,
where the nonlinear sigma model becomes strongly coupled.  
Here $f$ is the decay constant of
the pseudo-Goldstone boson and is necessarily of order a TeV.  The 
$SU(5)\to SO(5)$ symmetry breaking is generated at the scale 
$\Lambda$ by strong interactions similar to technicolor which act
only at the high scale $\Lambda$ and have no remnants at a TeV.  
The 14 Goldstone bosons remaining after
this symmetry breaking yield a physical doublet and a complex triplet
under $SU(2)$;  the remaining fields are eaten by a Higgs-like mechanism
when the intermediate symmetry is broken.
The one-loop contributions to the
quadratic divergences from the electroweak gauge sector are removed
by making use of the $SU(5)$ weakly gauged subgroup 
$[SU(2)\times U(1)]^2$.  The divergent contributions must then involve 
couplings of both groups and hence first appear at two-loop order.
This combination of gauge couplings breaks the global symmetries
and generates a small mass of order 100 GeV at the two-loop level for 
the scalar doublet field.  At one-loop order, the Higgs mass is not 
sensitive to the effective high scale $\Lambda$, and its 
mass is thus relatively stable assuming $\Lambda\sim 10$ TeV. 
The scalar triplet acquires a mass at the TeV scale from one-loop
gauge interactions.  A massive vector-like $SU(2)$ singlet $Q=+2/3$
quark is responsible for canceling the one-loop divergences
originating from the SM top-quark.

The physical spectrum of this model 
below a TeV is thus simply that of the SM with a single
light Higgs.  At the TeV scale four new gauge bosons 
$V_i$ (an electroweak triplet and singlet) appear, as well as the 
scalar triplet $\phi$, and a single vector-like quark $t^{'}$.  For the 
quadratically divergent corrections to cancel naturally,
and not be fine-tuned, the mass of the new quark is 
constrained to be \cite{hckn}
\begin{equation}
m_{t'}\lsim 2~{\rm TeV} \left[ {m_H\over 200~{\rm GeV}}\right]^2 \,,
\end{equation}
which in turn implies
\begin{equation}
f\lsim 1~{\rm TeV} \left[ {m_H\over 200~{\rm GeV}}\right]^2 \,.
\end{equation}
In addition, naturalness also implies
\begin{equation}
m_{V_i}\lsim 6~{\rm TeV} 
\left[ {m_H\over 200~{\rm GeV}}\right]^2\,, \quad\quad\quad
m_\phi\lsim 10~{\rm TeV}\,.
\end{equation}

The Littlest Higgs, based on a $SU(5)/SO(5)$ non-linear sigma model, 
is the simplest scenario of this type
in the sense that it introduces a minimal number of new fields.  Other 
models have been investigated \cite{others} and are 
based on the cosets $SU(k)/Sp(k)$, $SU(k)^2/SU(k)$, and 
$SU(k)^n/SU(k)^m$, where $k$ is the number of strongly interacting 
fermions at the scale $\Lambda$.  These
cases differ from the Littlest Higgs in that they require the
introduction of additional scalars, gauge bosons, and vector-like quarks
at the TeV scale, and sometimes extra light scalar fields
can arise at $\sim 100$ GeV.

In this paper, we examine the effects of the components of the Little
Higgs on precision electroweak observables and in direct searches for
new particles.  We concentrate on the minimal model here, the
Littlest Higgs \cite{hckn}, as it contains
all the essential ingredients and provides a representative 
example of these types of theories.  We would expect the more 
complicated scenarios listed above to be even further constrained by
experiment in the absence of any fine-tuning due to the presence of
the additional new particles at the TeV scale.
Consistency with the global precision electroweak data set is a tough 
barrier for any new model to pass, and we find that the Littlest Higgs
does not necessarily decouple quickly enough for most of the parameter
space.  A thorough examination of the parameter space
reveals small regions where precision data, alone, allows 
for scales in the theory to take on their natural
values, {\ie}, $f\lsim 2$ TeV. 
However, we find that the most stringent constraint arises from the
direct search for new gauge bosons in Drell-Yan production at the 
Tevatron \cite{CDF} and that this excludes these regions.
Together, the direct and indirect data samples
place the constraints $f\gsim 3.5 -4.0$ TeV, and $m_{t'}\gsim 7$ TeV, 
indicating that significant fine-tuning is required if this scenario is to 
work. 

The outline for the remainder of this paper is as follows:
in the next section we present the formalism for our analysis.
We start with a description of the parameters in this theory and then
give a derivation of the shifts that occur in the precision electroweak
observables.  In section 3, we present our numerical results.  We first
examine the bounds obtained from  
direct searches for new gauge bosons at the Tevatron.  We then
study the case where the new particles take on SM values for their coupling
strengths and perform a fit to the electroweak data set and
obtain tight constraints on the model.
We then vary these couplings within natural ranges and demonstrate
that there is a region of parameter space where the precision data
bounds are relaxed, but is still excluded by the Tevatron data.
The final section contains our conclusions.

\section{Formalism}

To begin our analysis, we follow the notation of 
Arkani-Hamed \etal {\cite {hckn}} and generate the masses of the
$[SU(2)\times U(1)]^2$ gauge bosons through 
the leading two-derivative kinetic term for the non-linear sigma model
\begin{equation}
{\cal L}={f^2\over {4}} Tr|D_\mu \Sigma|^2\,,
\end{equation}
where $f \sim 1$ TeV and the $\Sigma=exp(2i \Pi/f) \Sigma_0$ represents the 
Goldstone boson fields. The fourteen fields $\Pi$ include the SM 
weak iso-doublet $h$, an 
isotriplet $\phi$ as well as the true Goldstone bosons which are eaten in the 
breaking of the intermediate symmetry to the SM.  We choose the normalization
\begin{equation}
\Pi = \left( \begin{array}{ccc}
                   & h^\dagger/2 & \phi^\dagger/\sqrt 2   \\
              h/2  &             &  h^*/2                 \\
              \phi/\sqrt 2 & h^T/2 &                     
              \end{array}   \right) 
\end{equation}
such that the normalization of the kinetic term above is in agreement
with that in  earlier work on the non-linear sigma model \cite{kaplan}.
In addition to the usual Higgs doublet vev, the triplet $\phi$ also 
acquires an $SU(2)$ breaking vev.  We demonstrate in the Appendix that
including the contribution from the triplet vev to the precision
electroweak observables only strengthens our 
conclusions; here we adopt a conservative approach 
and neglect this piece in the following.  We note that
the presence of a scalar triplet is not a generic feature of 
Little Higgs models.  The covariant derivative of $\Sigma$ is given by
\begin{equation}
D\Sigma=\partial \Sigma -ig_kW^a_k(Q^a_k\Sigma+\Sigma Q^{aT}_k)-
ig_k' B_k(Y_k\Sigma+\Sigma Y^T_k)\,,
\end{equation}
where a sum over the index $k=1,2$ is understood and the `charge' ($Q^a$) and 
`hypercharge' ($Y$) matrices are 
given in Ref. {\cite {hckn}}. To proceed further in the calculation of the 
gauge boson masses it is useful to define the following combinations of the 
above $[SU(2)\times U(1)]_k$ gauge couplings: 
\begin{eqnarray}
g &=& \frac{g_1g_2}{\sqrt{g_1^2+g_2^2}} \,\, , \;\;\; 
g^{'} = \frac{g^{'}_1g^{'}_2}{\sqrt{g^{'2}_1+g^{'2}_2}} \,\, , \nonumber \\
g_t &=& \frac{g^{2}_1}{\sqrt{g_1^2+g_2^2}} \,\, , \;\;\; 
g^{'}_t = \frac{g^{'2}_1}{\sqrt{g^{'2}_1+g^{'2}_2}} \,\,.
\label{coupdef}
\end{eqnarray}
Inverting these relations yields
\begin{eqnarray}
g_1 &=& (g^2+g_t^2)^{1/2}\,, \nonumber \\
g_2 &=& {g\over {g_t}}(g^2+g_t^2)^{1/2}\,,
\end{eqnarray}
and similarly for $g_{1,2}'$. 
To zeroth-order in the Higgs vev, one linear combination of the gauge 
bosons obtains a mass at the TeV scale when the intermediate symmetry
is broken, while the second set remains massless. The massive 
neutral fields and their corresponding masses are given by 
\begin{eqnarray}
B_h &=& \frac{g^{'}_1B_1-g^{'}_2B_2}{\sqrt{g^{'2}_1+g^{'2}_2}} \,\, , \;\;\; 
m_{B_h}^2 = \frac{1}{10}\frac{\left(g^{'2}+g^{'2}_t\right)^2}{g^{'2}_t} f^2
\,\, , \nonumber \\
Z_h &=& \frac{g_1W^{3}_1-g_2W^{3}_2}{\sqrt{g^{2}_1+g^{2}_2}} \,\, , \;\;\; 
m_{Z_h}^2 = \frac{1}{2}\frac{\left(g^{2}+g^{2}_t\right)^2}{g^{2}_t} f^2\,\,.
\label{BhZhdef}
\end{eqnarray}
The corresponding charged isospin partner of the $Z_h$, $W_h^{\pm}$, obtains 
an identical mass:
\begin{equation}
W^{\pm}_h = \frac{g_1W^{\pm}_1-g_2W^{\pm}_2}{\sqrt{g_{1}^2+g_{2}^2}} 
\,\, , \;\;\; m_{W_h}^2 = \frac{1}{2}\frac{\left(g^{2}+g^{2}_t\right)^2}
{g^{2}_t} f^2\,\,.
\label{Whdef}
\end{equation}
The orthogonal linear combinations 
\begin{eqnarray}
B^{(0)} &=& \frac{g^{'}_2B_1+g^{'}_1B_2}{\sqrt{g^{'2}_1+g^{'2}_2}} \,\, ,
\nonumber \\ 
W^{3(0)} &=& \frac{g_2W^{3}_1+g_1W^{3}_2}{\sqrt{g^{2}_1+g^{2}_2}}\,\,,
\label{B0W0def}
\end{eqnarray}
together with the corresponding charged field, $W^{\pm (0)}$, 
remain massless at this 
order.  They will play the roles of the SM fields and acquire
masses through the usual Higgs mechanism which are proportional to 
the Higgs vev, $\nu$. 

In performing the comparison to precision 
electroweak and collider data, one must include 
the shifts in the SM-like gauge boson masses due to mixing with the 
heavier states, \ie, the Higgs vev induces a finite mixing between the
states $B_h,Z_h,W_h$ and the corresponding 
lighter fields. For the neutral fields, which 
form a $4\times 4$ mass matrix, this mixing is 
most easily examined via a power series expansion 
in the ratio of the Higgs vev relative 
to $f$.  For the charged fields, since the mass 
matrices are now only $2\times 2$, exact eigenvalues in a compact form can be  
easily found. The leading corrections to the masses 
of the lighter SM-like fields are of relative 
order $\delta^2= (\nu/f)^2 \sim 0.01$ so 
that, numerically,  
$\delta^4$ corrections can be safely neglected. When discussing the 
effects of the heavy gauge fields on precision measurements, we will 
consistently keep all terms of relative order $\delta^2$. Turning now to the 
light mass eigenstates, we define the zeroth-order 
weak mixing angle as follows:
\begin{equation}
c_{W}^{(0)} = \frac{g}{\sqrt{g^2+g^{'2}}}\,, \;\;\;\; 
s_{W}^{(0)} = \frac{g^{'}}{\sqrt{g^2+g^{'2}}} \,,
\label{sw0def}
\end{equation}
so that the physical $\gamma$ and $Z$ bosons become 
(to ${\cal O}\left(\delta^2\right)$)
\begin{eqnarray}
A &=& c_{W}^{(0)}B^{(0)}-s_{W}^{(0)}W^{3(0)} \,\, , \nonumber \\
Z &=& c_{W}^{(0)}W^{3(0)}+s_{W}^{(0)}B^{(0)}+\delta_{B_h} B_h
+\delta_{Z_h} Z_h \,\,,
\label{AZdef}
\end{eqnarray}
where $\delta_{B_h}\,, \delta_{Z_h}$ are defined below in Eq. 15.
For the physical $W^\pm$ bosons we obtain 
\begin{equation}
W^{\pm} = \frac{g_2}{\sqrt{g_{1}^2+g_{2}^2}}\left\{ 1-\frac{g_{1}^2
\left(g_{1}^2-g_{2}^2\right)}{4\left(g_{1}^2+g_{2}^2\right)^2} \delta^2 
\right\} W^{\pm}_{1} +
\frac{g_1}{\sqrt{g_{1}^2+g_{2}^2}}\left\{ 1+\frac{g_{2}^2
\left(g_{1}^2-g_{2}^2\right)}{4\left(g_{1}^2+g_{2}^2\right)^2} \delta^2 
\right\} W^{\pm}_{2}\,\,,
\label{Wdef}
\end{equation}
which can easily be written in terms of the unmixed 
fields $W^{\pm(0)}$ and $W_h^\pm$,
\begin{equation}
W^\pm= W^{\pm (0)}-{g_1g_2(g_1^2-g_2^2)\over {4(g_1^2+g_2^2)^2}}\delta^2
~W_h^\pm\,. 
\end{equation}
Note that the mixing between these states vanishes in the limit 
$g_1=g_2$ or, more specifically, when $g_t=g$. 
The physical $Z$ contains a small admixture
of the heavy fields proportional to $\delta_{B_h},\delta_{Z_h}$. These 
parameters, which define the mixing between the light SM-like $Z$ and the 
heavy neutral gauge bosons, are given to order $\delta^2$ by:
\begin{eqnarray}
\delta_{B_h} &=& -\frac{5}{4}\frac{g^{'}_t\left(g^{'2}_t-g^{'2}\right)
\sqrt{g^2+g^{'2}}}{\left(g^{'2}+g^{'2}_t\right)^2} \delta^2\,\, , \nonumber \\
\delta_{Z_h} &=& -\frac{1}{4}\frac{g_t\left(g^{2}_t-g^{2}\right)\sqrt{g^2
+g^{'2}}}{\left(g^{2}+g^{2}_t\right)^2}\delta^2\,\,.
\label{mixdef}
\end{eqnarray}
Again we note that these mixing terms vanish in the symmetric limit when 
$g_t=g$ and $g_t'=g'$. From this we see that {\it all} light-heavy gauge 
boson mixing vanishes in the symmetric case. This will have important 
consequences below. 
The masses of the physical $W^{\pm}$ and $Z$ bosons are then given
by (to ${\cal O}\left(\delta^2\right)$):
\begin{eqnarray}
M_{Z}^2 &=& m_{Z}^{2(0)} \left\{1-\delta^2 ~\frac{5\left(g_{t}^4+g^4\right)
\left(g_{t}^{'4}-g^{'2}g^{'2}_t+g^{'4}\right)+7g^2g_{t}^2\left(g^{'2}
+g^{'2}_t\right)^2-16g^2g^{'2}g_{t}^2g^{'2}_t}{6\left(g^2+g_{t}^2\right)^2
\left(g^{'2}+g^{'2}_{t}\right)^2} \right\} \,\, , \nonumber \\
M_{W}^2 &=& m_{W}^{2(0)} \left\{1-\delta^2 ~\frac{5g^4-2g^2g_{t}^2+5g_{t}^4}
{24\left(g^2+g_{t}^2\right)^2}\right\}\,\,,
\label{massdef}
\end{eqnarray}
where
\begin{equation}
m_{Z}^{2(0)} = \frac{1}{4}\left(g^2+g^{'2}\right)\nu^2 \,\, , \;\;\; 
m_{W}^{2(0)} = \frac{1}{4}g^2\nu^2
\label{zeromassdef}
\end{equation}
are the usual gauge boson masses that satisfy the tree-level SM relations: 
$m_Z^{(0)}c_W^{(0)}=m_W^{(0)}$.   

These shifts in the masses of the gauge fields 
contribute to the deviation of the $\rho$-parameter from its SM value.
These contributions alone yield
\begin{eqnarray}
\delta\rho_G & = & {\delta M_W^2\over m_W^{2(0)}} - 
{\delta M_Z^2\over m_Z^{2(0)}} \\
& = & \frac{5}{8}\left(\frac{g^{'2}_t-g^{'2}}
{g^{'2}_t+g^{'2}}\right)^2 \delta^2 \nonumber\,\,.  
\label{drhodef}
\end{eqnarray}
Note that as we might have anticipated 
the gauge contribution to $\delta \rho$ vanishes in the symmetric 
limit, \ie, when $g_t'=g'$. This is expected as in this limit the $W$ 
and $Z$ are just the usual SM fields. 
As we will discuss below there is an additional new contribution to 
$\delta \rho$ beyond that arising from 
the tree level shifts of the gauge boson masses described here. This is due 
to the mixing of the top quark with the new vectorlike isosinglet field, 
$t^{'}$. 
This contribution, which we denote by $\delta\rho_{top}$, arises from the 
new $t$ and $t'$ 
quark loops that do not appear in the SM. As we will see, these mixing 
induced  
contributions can be numerically significant in some regions of the parameter 
space and must be included. The reason for their significance 
is that such terms can be 
parametrically enhanced by powers of $m_t^2/M_W^2$ or $m_t'^2/M_W^2$ (though 
correspondingly suppressed by powers of the $t-t'$ mixing angle). 
Note that even though the additional isosinglet quark is vector-like, 
since it mixes with the top quark it does not fully decouple, \ie, it can lead 
to small but important corrections to $\delta \rho$ via mixing over some 
parameter space regions-particularly those where the $t-t'$ mixing angle 
is large. (Here we do not mean that as the $t'$ mass 
increases its contribution to $\delta \rho$ does not go to zero, 
\ie, the classical meaning of decoupling.) Thus the 
complete new contribution to the shift in the $\rho$-parameter
is the sum of $\delta\rho_{top}$ and $\delta\rho_G$. 

To proceed further with an analysis of precision data and collider 
constraints, we must examine how the SM fermions transform under the group 
$G_1\times 
G_2$ with $G_k=[SU(2)\times U(1)]_k$.  Here we take the fermions to transform 
non-trivially only under $G_1$. This choice is not unique and 
at least two other possibilities exist. The first alternative 
is when the fermions transform identically under both $G_1$ 
and $G_2$. This possibility is excluded here since it would require 
a doubling of 
the number of fermions. A second somewhat convoluted alternative would have 
the fermions transform under, say, $G_1$ as well as the $U(1)$ piece of $G_2$.
In this case, doubling of the number of fermions is not required, but
such a possibility would 
allow for an additional free parameter in the fermion couplings to the $Z$. 
It would seem that our choice is more natural and economical 
and avoids the issue 
of fermion doubling. However, we will make some particular notes 
in our numerical analysis 
below of the dependency of the choice of fermion couplings to the heavy 
hypercharge gauge boson.

The $t$ and $t'$ quarks contribute at one-loop to $\delta\rho$ as 
discussed above, as well as to $R_b$, the ratio of the b-quark to total
hadronic width of the $Z$.
In order to compute these contributions  
we must examine the $t-t'$ mass matrix which can be obtained from the 
Yukawa couplings provided in {\cite {hckn}}. The diagonalization of 
this matrix requires a bi-unitary transformation, which for real Yukawa
couplings implies that two separate rotations on the $t$ and $t'$ 
fields, $O_{L,R}$, must be performed. Here $O_{L,R}$ 
rotates the left(right)-handed components of the quark fields. 
It is important to note that the rotation $O_R$ 
is actually unphysical since both $t_R$ and $t'_R$ transform identically 
under $G_1$. $O_L$, however, is physical and leads to, \eg,  flavor violating
$Ztt'$ couplings. Since the mass matrix is a simple $2\times 2$ 
rotation, the matrix $O_L$ contains 
only a single parameter: a mixing angle $\theta_t$. Writing the ratio of the 
Yukawa couplings 
as $r=\lambda_2^2/\lambda_1^2$ and recalling $\delta=\nu/f$ one finds 
\begin{equation}
\tan (2\theta_t)={\sqrt {2} \delta\over {1+r-\delta^2/2}}\,\,,
\end{equation}
from which it follows that 
\begin{equation}
r=\sqrt {2} \delta/\tan (2\theta_t)-1+\delta^2/2\,.
\end{equation}
Since by definition $r \geq 0$, this immediately 
provides a bound on $\theta_t$. We find that 
\begin{equation}
\tan (\theta_t) \leq {\delta\over {\sqrt {2}}}\,.
\end{equation}
Since $\delta$ is relatively small this greatly restricts the range of 
$\theta_t$. The eigenvalues of the $t-t'$ mass matrix are now 
easily obtained with the smaller one being 
identified with the physical top quark. Given $\delta, \theta_t$ and 
the fixed experimental value of $m_t=174.3\pm 5.1$ GeV \cite{pdg},  
the mass of the $t'$, $m_{t'}$, is now completely determined:
\begin{equation}
{m_{t'}^2\over {m_t^2}}={{1+r+\delta^2/2+\sqrt {(1+r+\delta^2/2)^2-2r\delta^2}}
\over {1+r+\delta^2/2-\sqrt {(1+r+\delta^2/2)^2-2r\delta^2}}}\,.
\end{equation}
$m^2_{t'}$ is then given by the numerator in the above equation times
a factor of $\lambda_1^2 f^2$.

For a given set of values for the parameters $(\delta,\theta_t)$, the
contribution to the shift in the $\rho$-parameter from the fermion
sector can now 
be fully calculated. (For simplicity we set $V_{tb}=1$ in what follows.) 
The couplings of the $t$ and $t'$ to the $W$ and $Z$ 
fields are altered by the presence of the mixing and are represented by the 
parameter $\theta_t$, \eg, ordinarily the 
isosinglet $t'$ would not couple to the $W$. We find $W\bar t b\sim c_t$ 
while $W\bar t' b\sim s_t$ 
where $c_t[s_t]=\cos (\theta_t)[\sin (\theta_t)]$. 
Note that to this order in $\delta$ we can 
neglect the effects of gauge boson mixing on the couplings of the $W$ and 
$Z$.  For the $Z$, the right-handed couplings of the $t$ and $t'$ 
are not modified but the 
left-handed couplings are now given by 
\begin{eqnarray}
g_{L}^{tt} &=& \sqrt{\frac{8G_fM_{Z}^2}{\sqrt{2}}} \left( c_t^2/2-2x_w/3 
\right)\,\, ,\nonumber \\
g_{L}^{t^{'}t^{'}} &=&  \sqrt{\frac{8G_fM_{Z}^2}{\sqrt{2}}} \left(s_t^2/2
-2x_w/3\right) \,\, ,  \nonumber \\
g_{L}^{tt^{'}} &=& g_{L}^{t^{'}t}= \sqrt{\frac{8G_fM_{Z}^2}{\sqrt{2}}} 
\left(c_t s_t/2\right) \,\, , \nonumber \\
\end{eqnarray}
where $x_w=\sin^2\theta_W$ can be taken to be the on-shell value to this order 
in $\delta$. Note the presence of the 
flavor-changing coupling that we alluded to above. 
There are now five graphs which contribute to the vacuum polarization
of the $Z$ boson and two graphs for the 
$W$; the usual SM contribution must be subtracted from these new
contributions to $\delta\rho_{top}$. To 
proceed let us write the general couplings of a pair of fermions to a gauge 
boson as
\begin{equation}
{\cal L}= \bar f_1\gamma_\mu (v+a\gamma_5)f_2 X^\mu +h.c.
\end{equation}
In this language, we now find that 
\begin{equation}
\delta \rho={3G_f\over {2\sqrt 2 \pi^2}} (\delta_W-\delta_Z c_W^2)
\,,
\end{equation}
where $c_W=\cos \theta_W$, which, 
again to this order, can be taken to be the on-shell SM value, and 
\begin{eqnarray}
\delta_X &=& 2[2m_1m_2(v^2-a^2)-(m_1^2+m_2^2)(v^2+a^2)]\log(m_1m_2/\mu^2)
-2m_1m_2(v^2-a^2) \\
 &+& (m_2^2-m_1^2)^{-1}[2m_1m_2(m_1^2+m_2^2)(v^2-a^2)-(m_1^4+m_2^4)(v^2+a^2)]
\log(m_2^2/m_1^2)\,, \nonumber
\end{eqnarray}
with $m_{1,2}$ being the relevant fermion masses. 
Here, $\mu$ is an arbitrary renormalization scale that cancels from the 
final expression after all the individual diagrams are summed over. 
For the $Z$ boson vacuum polarization diagrams, the
intermediate states which contribute are $\bar tt,\bar t't',\bar tt',
\bar t't$ and $\bar bb$, while for the $W$ they are $\bar t b$ and 
$\bar t' b$, and their couplings are as given above. 
Using this expression we can now sum the new contributions and subtract 
those of the SM thus obtaining the contribution to the shift in the
$\rho$-parameter from the fermion sector of this model.

We now have the necessary ingredients to calculate the shifts in the
precision electroweak observables.  We perform our calculations in 
the on-shell renormalization scheme and begin by examining the effects
related to the definition of $G_f$ in this model.  In the on-shell 
renormalization scheme, the $W^{\pm}$ boson mass is defined   
through $G_f$, which is derived from muon decay, and $\alpha$, the fine 
structure constant.  Note that in the on-shell scheme, $G_f$ is employed
as an input parameter.  Several effects 
now modify this relation: (1) there are contributions from the exchange 
of the new gauge boson $W^{\pm}_h$ in $\mu$-decay; 
(2) the coupling of fermions to the 
physical $W^{\pm}$ are shifted from the  
expression in Eq.~\ref{coupdef} due to mixing with the $W^{\pm}_h$ as 
presented in Eq.~\ref{Wdef}:
\begin{equation}
g_{physical} \equiv g_W = g\left(1-\Delta g\right) \,\, , \;\;\; 
\Delta g = \frac{1}{4}\frac{g_{1}^2\left(g_{1}^2-g_{2}^2\right)}
{\left(g_{1}^2+g_{2}^2\right)^2} \delta^2 \,\, ;
\end{equation}
(3) the expression for $g$ in terms of $\alpha$ becomes 
$g^2 = 4\pi\alpha /s_{W}^{(0)2}$, where $s_{W}^{(0)}$ is defined in 
Eq.~\ref{sw0def}; this must be written in terms of the physical 
$W^{\pm}$ and $Z$ masses.  After implementing these changes, we arrive 
at the following equation relating $M_W$ to $G_f$ and $\alpha$
\begin{equation}
\frac{G_f}{\sqrt{2}} = \frac{\pi\alpha\left(1-\Delta g\right)^2}
{2M_{W}^2\left(1-M_{W}^2 /M_{Z}^2\right)\left(1+c_{W}^2 \delta\rho 
/s_{W}^2\right)} + \frac{g_{t}^2}{8m_{W_h}^2}\,\,.
\label{Gfdef}
\end{equation}
Here $\delta \rho=\delta \rho_G+\delta \rho_{top}$.
Solving this for $M_W$ to ${\cal O}\left(\delta^2\right)$, we find the 
following shift in the $W^{\pm}$ mass from its predicted SM value,
\begin{equation}
\delta M_{W}^2 = \frac{(M_{W}^2)_{SM} s_{W}^2}{1-2s_{W}^2} \left\{
\frac{c_{W}^2}{s_{W}^2}\delta\rho - \frac{1}{2}\frac{g^2g^{2}_{t}}
{\left(g^2+g^{2}_{t}\right)^2} \delta^2 \right\}\,.
\label{deltamw}
\end{equation}
We define $(M_W)_{SM}$ as the mass of the 
$W$ boson given by the SM when the Higgs boson mass, $m_H$, as well as  
$M_Z$, $G_f$, $\alpha(M_Z^2)$, $\alpha_s(M_Z^2)$ and $m_t$ are used as input, 
and when all of the usual 
SM radiative corrections are incorporated,
which we do here by using ZFITTER{\cite {zfitter}}. In 
evaluating this expression to order $\delta^2$, we let can let both 
$c_W$ and $s_W$ take their on-shell values. 

The on-shell expression for $s_{W}^2$  
which enters the $Z$ couplings is that defined 
in Eq.~\ref{sw0def}, and can be written as
\begin{equation}
s_{W}^{(0)2} = 1-\frac{m_{W}^{2(0)}}{m_{Z}^{2(0)}}\,\,.
\label{swosdef}
\end{equation}
In terms of the physical $Z$ mass and the SM prediction for 
the $W^{\pm}$ mass, we find the following shift in the on-shell definition 
of $s_{W}^{2}$: $s_{W}^{(0)2}\to s_{W}^{2}=s_{W}^{(0)2}+\delta s_{W}^{2}$ 
where 
\begin{equation}
\delta s_{W}^{2} = c_{W}^2 \delta\rho -\frac{\delta M_{W}^2}{M_{Z}^2}\,.
\label{dswosdef}
\end{equation}

At this point we next examine the interactions of the $Z$ with the various 
fermion fields. 
The SM expressions for a fermion's vector and axial-vector couplings 
to the $Z$ are 
\begin{equation}
g^{f}_v = \sqrt{\frac{8G_fM_{Z}^2}{\sqrt{2}}} \left(\frac{T^{f}_3}{2} 
-Q^f s_{W}^2\right) \,\, , \;\;\; 
g^{f}_a = \sqrt{\frac{8G_fM_{Z}^2}{\sqrt{2}}} 
\left(-\frac{T^{f}_3}{2}\right)\,\,. 
\label{SMcoupdef}
\end{equation}
These couplings are now modified in several ways: ($i$) the definition of 
$G_f$ has been changed as described above, 
($ii$) the couplings of the lighter SM-like $Z$ 
are shifted via the modification in $s_W^2$, ($iii$) the physical $Z$ now has 
additional components from the heavy gauge fields. When these effects are 
combined the resulting shifts in $g_v^f$ and $g_a^f$ are given by
\begin{eqnarray}
\delta g^{f}_v &=& \sqrt{\frac{8G_fM_{Z}^2}{\sqrt{2}}} 
\left\{\left( \frac{1}{2}
\delta\rho -\frac{1}{4}\frac{g^2g_{t}^2}{\left(g^2+g_{t}^2\right)^2}\delta^2
\right)\left(\frac{T^{f}_3}{2}-Q^f s_{W}^2\right) -Q^f \delta s_{W}^2 
\right\} \nonumber \\ & & + 
\frac{T^{f}_3}{2}\left(g_t \delta_{Z_h} +g^{'}_t \delta_{B_h}\right)
-g^{'}_t Q^f \delta_{B_h} \,\, , \nonumber \\ 
\delta g^{f}_a &=& \sqrt{\frac{8G_fM_{Z}^2}{\sqrt{2}}} 
\left\{\left( \frac{1}{2}
\delta\rho -\frac{1}{4}\frac{g^2g_{t}^2}{\left(g^2+g_{t}^2\right)^2}\delta^2
\right)\left(-\frac{T^{f}_3}{2}\right)\right\} 
- \frac{T^{f}_3}{2}\left(g_t \delta_{Z_h} +g^{'}_t \delta_{B_h}\right)\,\,.
\label{dgvgadef}
\end{eqnarray}
For $b$ quarks, there is an additional shift arising from vertex corrections
containing the extra heavy isosinglet quark as well as those involving the 
top since its couplings have been modified as discussed above. 
Expressing this as a shift 
in the left and right-handed couplings, with $g^{b}_v=(g^{b}_L+g^{b}_R)/2$ and 
$g^{b}_a = (g^{b}_R-g^{b}_L)/2$, and keeping only terms which
are quadratically enhanced by the 
top quark or heavy isosinglet quark masses, we find
\begin{eqnarray}
\delta g^{b}_L &=& \frac{G_f}{4\pi^2\sqrt{2}} \, \bigg[ \,
\sum_{i,j=t,t^{'}} m_im_jx_ix_j\left\{g_L^{ij} I_1\left(m_i,m_j\right)
-g^{t}_R I_2\left(m_i,m_j\right) 
\right. \nonumber \\ 
& & \left.  -\frac{g_z\left(s_{W}^2-c_{W}^2\right)}{2}\delta^{i}_{j} 
I_{3}\left(m_i\right) +g_{L}^{b}\delta^{i}_{j}
I_{4}\left(m_i\right) \right\} 
- m_{t}^2 \left\{ g_{L}^{t} I_1\left(m_t,m_t\right) \right. 
\nonumber \\ & & \left. 
 -g_{R}^{t}I_2\left(m_t,m_t\right)-\frac{g_z\left(s_{W}^2
-c_{W}^2\right)}{2}I_{3}\left(m_t\right) +g_{L}^{b}
I_{4}\left(m_t\right) \right\} \bigg] \,\, ,
\label{loopdef}
\end{eqnarray}
where $g_z = \sqrt{\frac{8G_fM_{Z}^2}{\sqrt{2}}} $, and the functions 
$I_k$ are defined in the Appendix.  The $x_i$ denote the 
mixing angles which arise when deriving the charged Goldstone coupling 
to the physical $t$ and $t^{'}$ states; we find $x_t = {\rm cos}\left(
\theta_t\right)$, $x_{t'} = -{\rm sin}\left(\theta_t\right)$.  We have 
explicitly subtracted off the SM contribution in Eq.~\ref{loopdef}.  This 
shift requires no renormalization, and is therefore a prediction of 
this model. We find that the right-handed coupling $g^{b}_R$ receives no 
contributions which are quadratically enhanced by the masses
of the $t$ or $t^{'}$.

Using the above expressions, we can derive the shifts in the precision 
EW observables.  We examine the following approximately uncorrelated set of 
of observables: $M_{W}$, $\Gamma_l$, $s_{W,eff}^{2,lep}$, $N_\nu$, $R_b$, 
$A_b$, $R_c$, and $A_c$.  Here $\Gamma_l$ is the total leptonic width
of the $Z$, 
$s_{W,eff}^{2,lep}$ is the effective leptonic weak mixing angle on the $Z$ 
pole, $N_\nu$ 
is the `number of neutrinos' (defined through the invisible width of the $Z$), 
$R_b$ and $R_c$ are respectively the ratios of bottom and charm quark widths 
over the total hadronic width, and $A_b$ and $A_c$ are respectively the 
polarization asymmetries for $b$ and $c$ quarks.  We will also dicuss the 
deviation induced in the 
NuTeV measurement of the on-shell weak mixing angle~\cite{Zeller:2001hh}.  
Once the various coupling shifts are determined, it is rather easy 
to obtain the contributions to these observables.  We find:
\begin{eqnarray}
\delta \Gamma_f &=& N^{f}_c \frac{M_{Z}}{6\pi^2}\left(g^{f}_{v} 
\delta g^{f}_{v}
+ g^{f}_{a} \delta g^{f}_{a} \right) \,\, , \nonumber \\
\delta s_{W,eff}^{2,lep} &=& \frac{g^{l}_{a} \delta g^{l}_{v} 
-g^{l}_{v} \delta g^{l}_{a}}{4 g_{a}^{l2}} \,\, , \nonumber \\
\delta N_\nu &=& 3 \left\{ 2 \left(\frac{g^{\nu}_{v}\delta g^{\nu}_{v}
+g^{\nu}_{a}\delta g^{\nu}_{a}}{g^{\nu 2}_{v}+g^{\nu 2}_{a}} \right) 
-1 \right\} \,\, , \nonumber \\
\delta R_q &=& \frac{\Gamma_{had}\delta\Gamma_q - \Gamma_q 
\delta\Gamma_{had}}{\Gamma_{had}^2}  \,\, , \nonumber \\
\delta A_q &=& 2 \frac{g^{q2}_{v}-g^{q2}_{a}}{\left(g^{q2}_{v}+g^{q2}_{a}
\right)^2} \left(g^{q}_{v}\delta g^{q}_{a} - g^{q}_{a}\delta g^{q}_{v}
\right) \,\, .
\label{EWshifts}
\end{eqnarray}
The NuTeV experiment measures the following ratio of neutral and charged 
current cross sections for both neutrino and anti-neutrino beams 
scattering off an isoscalar target: 
\begin{equation}
R_{-} = \frac{\sigma^{\nu}_{NC} - \sigma^{\bar{\nu}}_{NC}}
{\sigma^{\nu}_{CC} - \sigma^{\bar{\nu}}_{CC}} = \frac{1}{2}-s_{W}^{2}
\,\, ,
\label{PWrelation}
\end{equation}
where in obtaining the second equality we have included the tree-level SM 
prediction for $R_{-}$.  In the present model, 
the charged current contribution, 
which now includes the exchange of both $W^{\pm}$ and $W^{\pm}_h$, can be 
expressed entirely in terms of $G_f$.  This is unchanged in this model since 
$G_f$ is taken as an input parameter 
in the electroweak fit.  The neutral current piece is 
modified by both the shifted $Z\bar{f}f$ couplings discussed above and the 
exchange of the new heavy gauge bosons $B_h$ and $Z_h$.  Given the masses 
and couplings of these gauge bosons, it is straightforward to 
compute the required diagrams and find the shift in $R_{-}$.  The masses 
of $B_h$ and $Z_h$ are given in Eq.~\ref{BhZhdef}; writing the interaction 
between these states and the 
fermion fields as $(g^{f,X}_v + g^{f,X}_a \gamma_5)$, 
we find 
\begin{eqnarray}
g^{f,Z}_v &=& g_t T^{f}_{3}/2 \,\, , \;\;\; 
g^{f,Z}_a = -g_t T^{f}_{3}/2 \,\, , \nonumber \\
g^{f,B}_v &=& g^{'}_{t}\left(T^{f}_{3}-2Q^{f}\right)/2 \,\, , \;\;\; 
g^{f,B}_a = -g^{'}_t T^{f}_{3}/2 \,\, .
\label{ZhBhcoups}
\end{eqnarray}
The exact expression for $R_-$ is not very enlightening so we do not include 
it here.  $R_{-}$ can be generally written in the form $\rho\left(1/2
-s_{W,N}^{2,os}
\right)$; in our analysis we will set $\rho=1$ and interpret the 
deviation as a shift in the on-shell weak mixing angle $s_{W,N}^{2,os}$.

From the discussion above it is clear that there are only four 
independent parameters that we need to consider in our numerical fits: 
$\delta$, $\sin \theta_t$ 
and the ratios of couplings $g_t/g$ and $g_t'/g'$. All of the shifts in the 
electroweak observables as well as the 
predicted masses for the new gauge bosons 
can be written in terms of these few parameters. 

\section{Results}

In this section we present our numerical
analysis of the experimental constraints 
on the Littlest Higgs, including those from both direct searches and 
precision measurements.  We attempt to identify throughout the 
features of our analysis that are independent of the fermion couplings 
to the hypercharge bosons.

\begin{figure}[htbp]
\centerline{
\psfig{figure=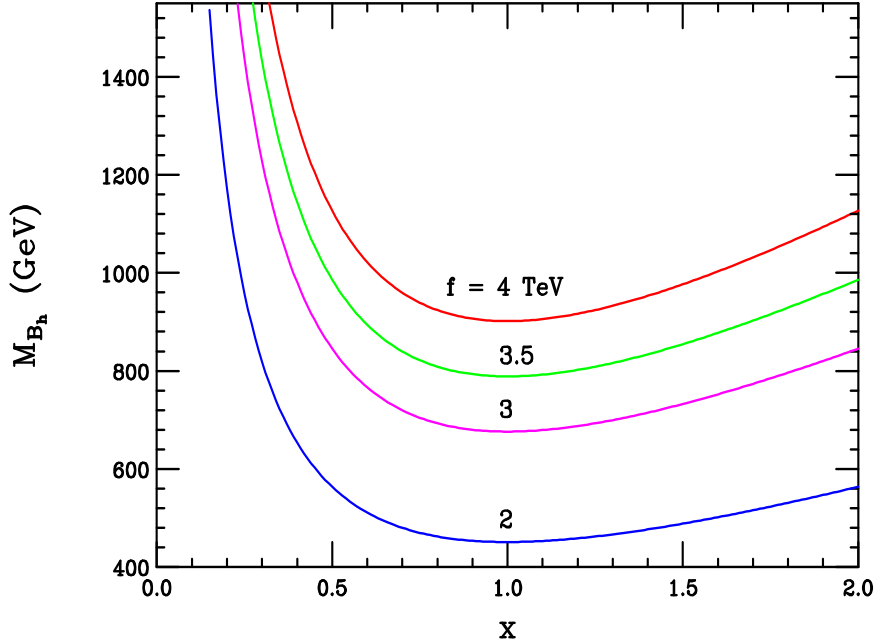,height=8.5cm,width=11.5cm,angle=0}}
\caption{The theoretical predictions for the 
mass of the gauge boson $B_h$ as a function of 
$x=g^{'}_{t}/g^{'}$, the ratio of its coupling relative to the corresponding 
SM coupling, for several values of the scale $f$.}
\label{mbmass}
\end{figure}

\begin{figure}[t]
\centerline{
\psfig{figure=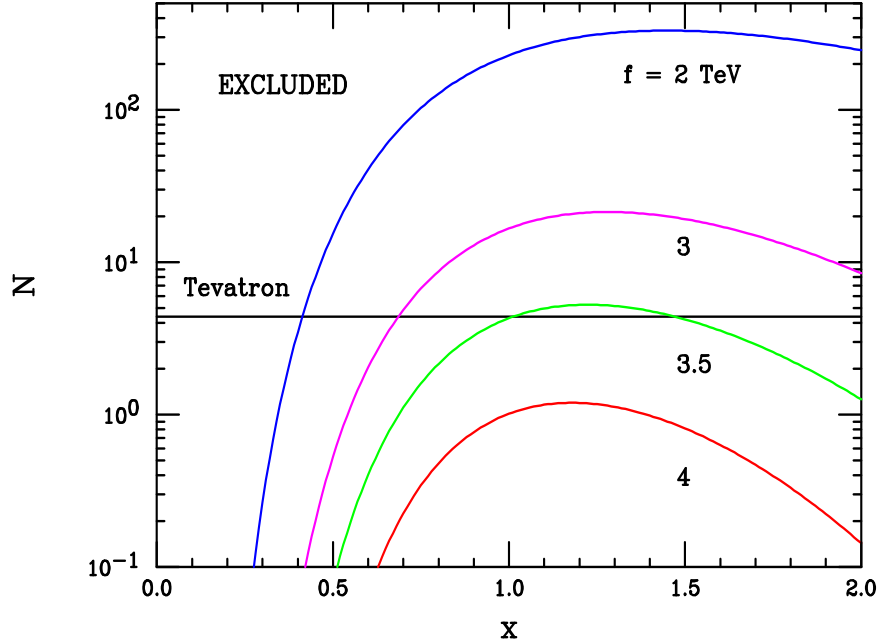,height=8.5cm,width=11.5cm,angle=0}}
\caption{The expected number of Drell-Yan events from the production
of the $B_h$ boson during Tevatron Run I 
(110 ${\rm pb}^{-1}$) as a function of $x=g^{'}_{t}/g^{'}$, the ratio of its 
coupling relative to 
the corresponding SM coupling, for several values of $f$.  The
horizontal line represents the 
95\% CL bound from CDF \cite{CDF} for this
mechanism; the parameter region above this line 
is excluded.}
\label{mbbound}
\end{figure}

The six new TeV-scale states predicted in this model are the four 
new gauge bosons, $B_h$, $Z_h$, and $W^{\pm}_h$, the isosinglet quark 
$t^{'}$, and the isotriplet $\phi$.  Limits on the masses and couplings 
of such particles have been set by direct searches at high-energy 
colliders such as the Tevatron.  The strongest constraints arise from
the lack of observation for the production of $B_h$. 
This can be understood from examining Eq.~\ref{BhZhdef}; the factor of 
$1/\sqrt{10}$ in its mass, which arises from the normalization of 
the hypercharge generator defined in~\cite{hckn}, and the fact that 
$g^{'} \approx .3 < g$ in the SM, show that it is predicted to be
quite light relative to the scale $f$.  
We present the theoretical predictions for $m_{B_h}$ as a function of 
the coupling ratio $x=g^{'}_{t}/g^{'}$ in Fig.~\ref{mbmass} for several 
values of $f$; $m_{B_h}$ can become as small as $~400$ GeV for $f$ as 
large as 2 TeV and $x \approx 1$.  The 95\% CL bounds resulting from direct 
searches at the 
Tevatron are shown in Fig.~\ref{mbbound}.  As expected, the strongest 
constraints are for $x \approx 1$, where $m_{B_h}$ reaches its minimum
value; in this region, the data bounds
$f \gsim 3.5 $ TeV.  The constraints weaken 
when $x \neq 1$.  However, as we will see later, $x \approx 1$ is the 
parameter choice preferred by the precision EW data; the combination 
of the direct and indirect 
constraints will require $f \gsim 3.5$ TeV for all $x$.  We 
summarize the constraints on $f$ arising from searches for 
$B_h$, $W^{\pm}_h$, and $Z_{h}$ production at the Tevatron Run I 
in Table~\ref{BhWhZhbounds}; as claimed above, the most 
significant limits are obtained from $B_h$ production.  We note that the 
constraints from $W^{\pm}_h$ and $Z_{h}$ production 
depend only on the coupling of the fermions to 
the SU$(2)_1$ gauge bosons, and not on their coupling to the hypercharge 
bosons.

\begin{table}[ht]
\begin{center}
\begin{tabular}{|c|c|c|c|}
\hline
 $x$ & $f \,\, (B_h)$ &  $f \,\, (W^{\pm}_h)$ & $f \,\, (Z_h)$ \\ \hline\hline
1 &  3.48 TeV & .85 TeV & .81 TeV \\ \hline
1/2 &  2.33 & .50 & .53 \\ \hline
2 &   3.23 & .65 & .75 \\ \hline\hline
\end{tabular}
\caption{$95\%$ CL lower bounds on $f$ for specific values of $x$ 
arising from Tevatron limits on $B_h$, 
$W^{\pm}_h$, and $Z_{h}$ production; for $B_h$, $x=g^{'}_{t}/g^{'}$, 
while for $W^{\pm}_h$ and $Z_h$, $x=g_{t}/g$.}
\label{BhWhZhbounds}
\end{center}
\end{table}

It is interesting to imagine how these constraints would evolve if the 
Tevatron Run II fails to find a signal for new gauge boson production in 
Drell-Yan collisions. What ranges of $x$ would now be allowed as $f$ is 
varied? The answer to this question can be found in Table~\ref{TevatronII}. 
We see that the sensitivity to $f$ would be extended by about $30\%$ in 
comparison to Run I which would most likely exclude the possibility that 
$f\leq 5$ TeV.

\begin{table}[ht]
\begin{center}
\begin{tabular}{|c|c|c|c|c|c|c|c|c|}
\hline
$f$ &   2 &   2.5 &   3 &   3.5 &   4 &   4.5   & 5 &   5.5 \\ \hline
$x$ &$\leq 0.24$&$\leq 0.26$&$\leq 0.30$&$\leq 0.43$&$\leq 0.51$&$\leq 0.61$
&$0.74-1.68$& all \\\hline\hline
\end{tabular}
\caption{Estimated allowed ranges of $x$ for different values of 
$f$ in TeV following 
a null search for new gauge bosons at the Tevatron Run II assuming an 
integrated luminosity of $10\, fb^{-1}$. For $f=5$ TeV, the {\it disallowed} 
range is shown.}
\label{TevatronII}
\end{center}
\end{table}

\vspace*{-1.0cm}
\noindent

\vspace*{-1.0cm}
\noindent

We now analyze the constraints arising from EW precision data.  The new  
parameters introduced in this model are the scale $f$ (or the ratio 
$\delta = \nu /f$), the couplings $g_t$ and 
$h_t$, and the top quark sector mixing angle $\theta_t$, which enters 
both $\delta\rho$ and $Z$-pole 
$b$-quark observables through the loop corrections 
discussed in the previous section.  Rather than perform a combined fit 
to all four quantities, we study a series of two-dimensional $\chi^2$ fits 
for various 
combinations of parameters; this is sufficient to illustrate the 
essential physics.  In deriving our constraints we follow the analysis 
of~\cite{Rizzo:1999br}; we calibrate our fits by first varying 
the Higgs boson mass until we find 
the $\chi^2$ value that reproduces the 95\% CL upper bound on $m_H$ 
presented in ~\cite{ewfit}.  We then use this reference $\chi^2$ to 
determine whether a given set of model parameters provides a good fit to the 
EW data.  We say that a set of parameters is disallowed if the resulting
$\chi^2$ exceeds this reference value.  We will first 
consider the observables $m_W$, $\Gamma_l$, $s_{W,eff}^{2,lep}$, $N_\nu$, 
$R_b$, $A_b$, $R_c$, and $A_c$, and later will include $s_{W,N}^{2,os}$ 
as determined from NuTeV.  The error correlation between these 
measurements is quite small~\cite{ewfit}, and will be neglected in our 
analysis.  We use 
ZFITTER~\cite{zfitter} to derive the SM predictions for these quantities.

\begin{figure}[htbp]
\centerline{\psfig{figure=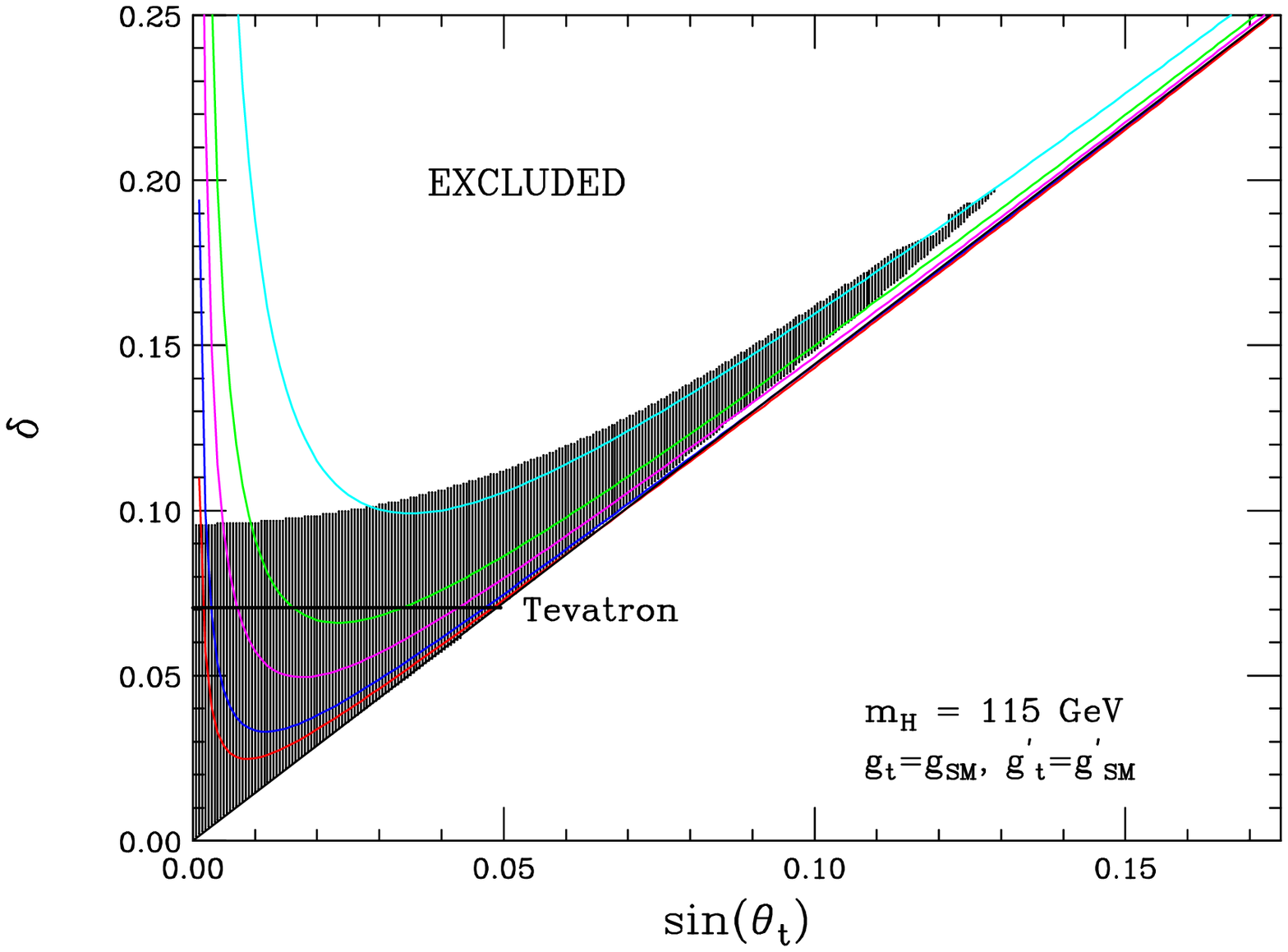,height=8.5cm,width=11.5cm,angle=0}}
\vspace*{1.0cm}
\centerline{\psfig{figure=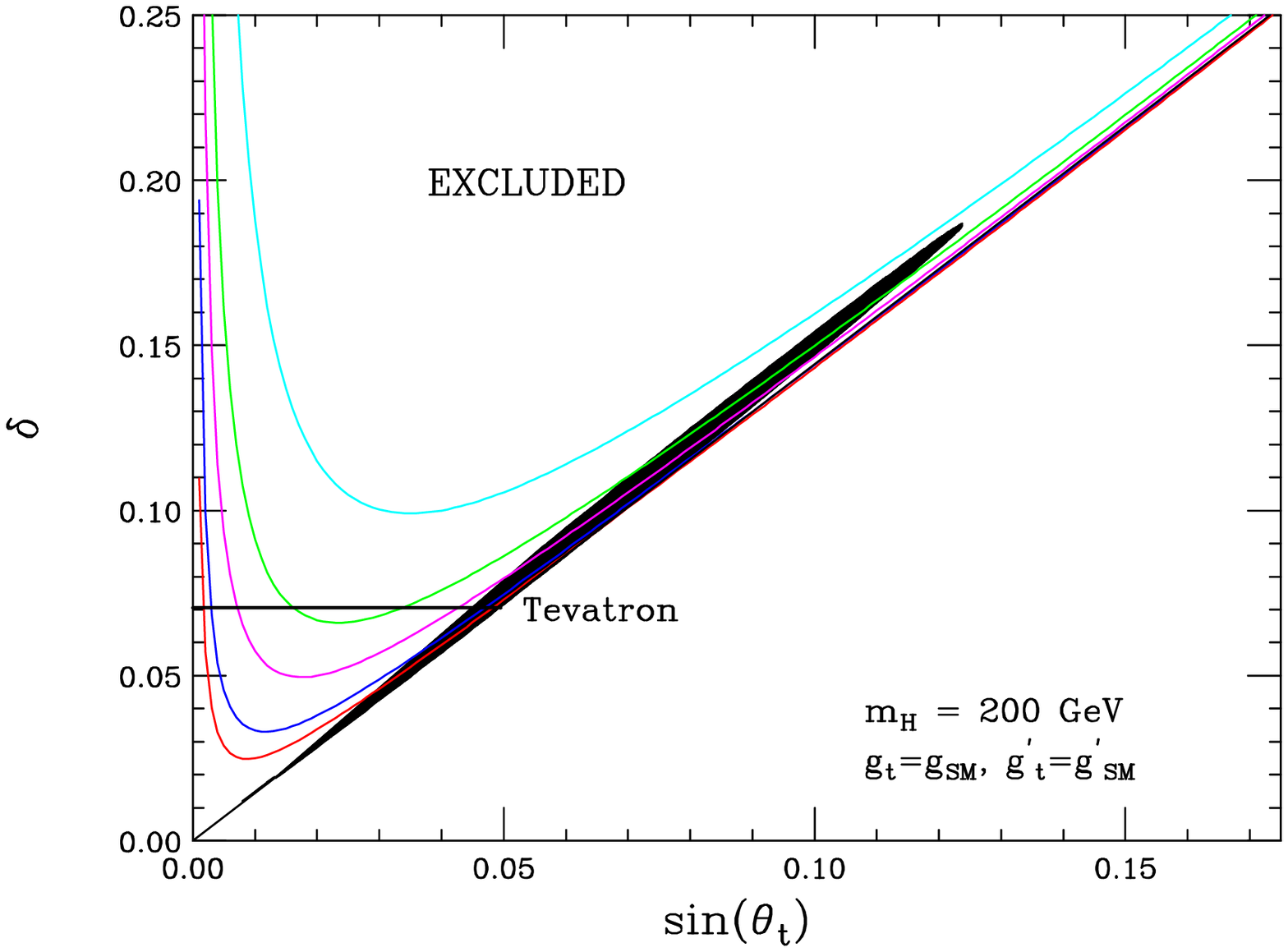,height=8.5cm,width=11.5cm,angle=0}}
\caption{Fit to the EW precision data varying ${\rm sin}
\left(\theta_t\right)$ and $\delta$, for $m_H=115$ GeV (upper) and 
$m_H = 200$ GeV (lower).  The diagonal line indicates the bound 
$\delta/\sqrt{2} \leq {\rm tan}\left(\theta_t\right)$, the horizontal line denotes 
the 95\% CL bound from $B_h$ production at the Tevatron.  The series of 
curved lines indicates the $t^{'}$ mass $m_{t^{'}}$ as a function of 
$\delta$ and ${\rm sin}\left(\theta_t\right)$; from top to bottom, they 
represent $m_{t^{'}} = 5$, 7.5, 10, 15, and 20 TeV.  The shaded regions are 
allowed by the EW fit.  For the remainder of our discussion, we label
$g$ and $g'$ as $g_{SM}$ and $g'_{SM}$, respectively, in the figures.}
\label{higgsvar}
\end{figure}

We first set $g_t=g$ and $g^{'}_{t} = g^{'}$, and vary both ${\rm sin}
\left(\theta_t\right)$ and $\delta$.  The results of this fit for both 
$m_H=115$ GeV and $m_H = 200$ GeV are shown in Fig.~\ref{higgsvar}; the 
shaded region is allowed by the EW fit at the 95\% CL, while the remainder 
is excluded.  Although 
$m_H = 200$ GeV is allowed in this model, unlike in the SM where 
$m_H < 193$ GeV at the 95\% CL, only a small sliver near the boundary 
$\delta/\sqrt{2} = {\rm tan}\left(\theta_t\right)$ satisfies the EW precision 
constraints.  This region persists but shrinks for larger values of $m_H$.  
The tail along this boundary for large values of 
$\delta$ is primarily due to 
the $t$ and $t^{'}$ mixing 
contributions to $\delta\rho$.  We see that the Tevatron 
constraints are quite significant here, as 
they completely exclude this region.  
Included in these figures are curves representing the predictions for 
several $t^{'}$ masses as 
functions of $\delta$ and ${\rm sin}\left(\theta_t\right)$ from
Eq. 22.  We see that the Tevatron limit requires that $m_{t^{'}} > 7$ TeV, 
which is significantly larger 
than the naturalness bound $m_{t^{'}} < 2$ TeV. This is beyond the 
range for direct discovery at the LHC. 

\vspace*{-0.8cm}
\noindent

\begin{figure}[htbp]
\centerline{
\psfig{figure=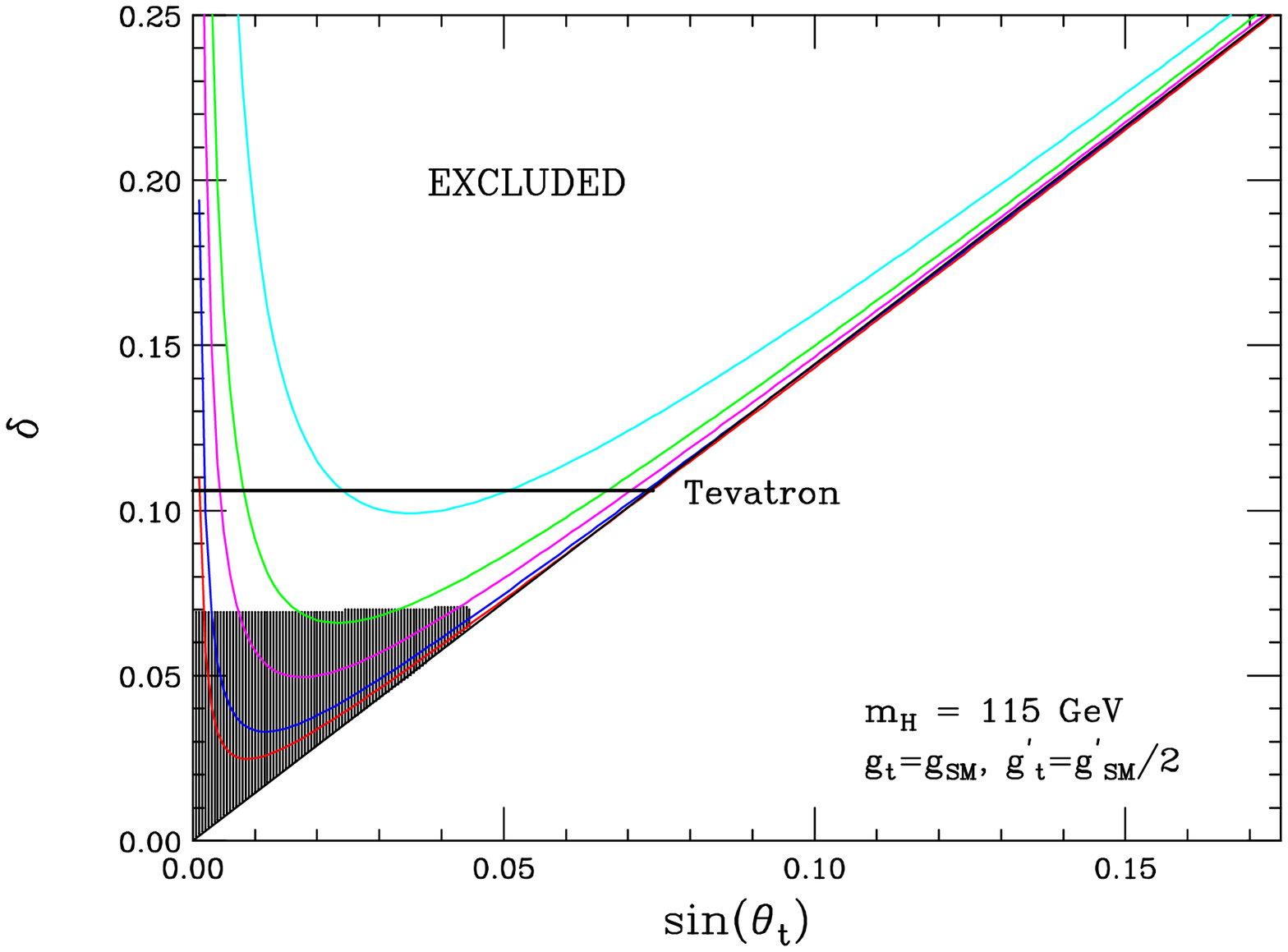,height=6.9cm,width=9.6cm,angle=0}
\psfig{figure=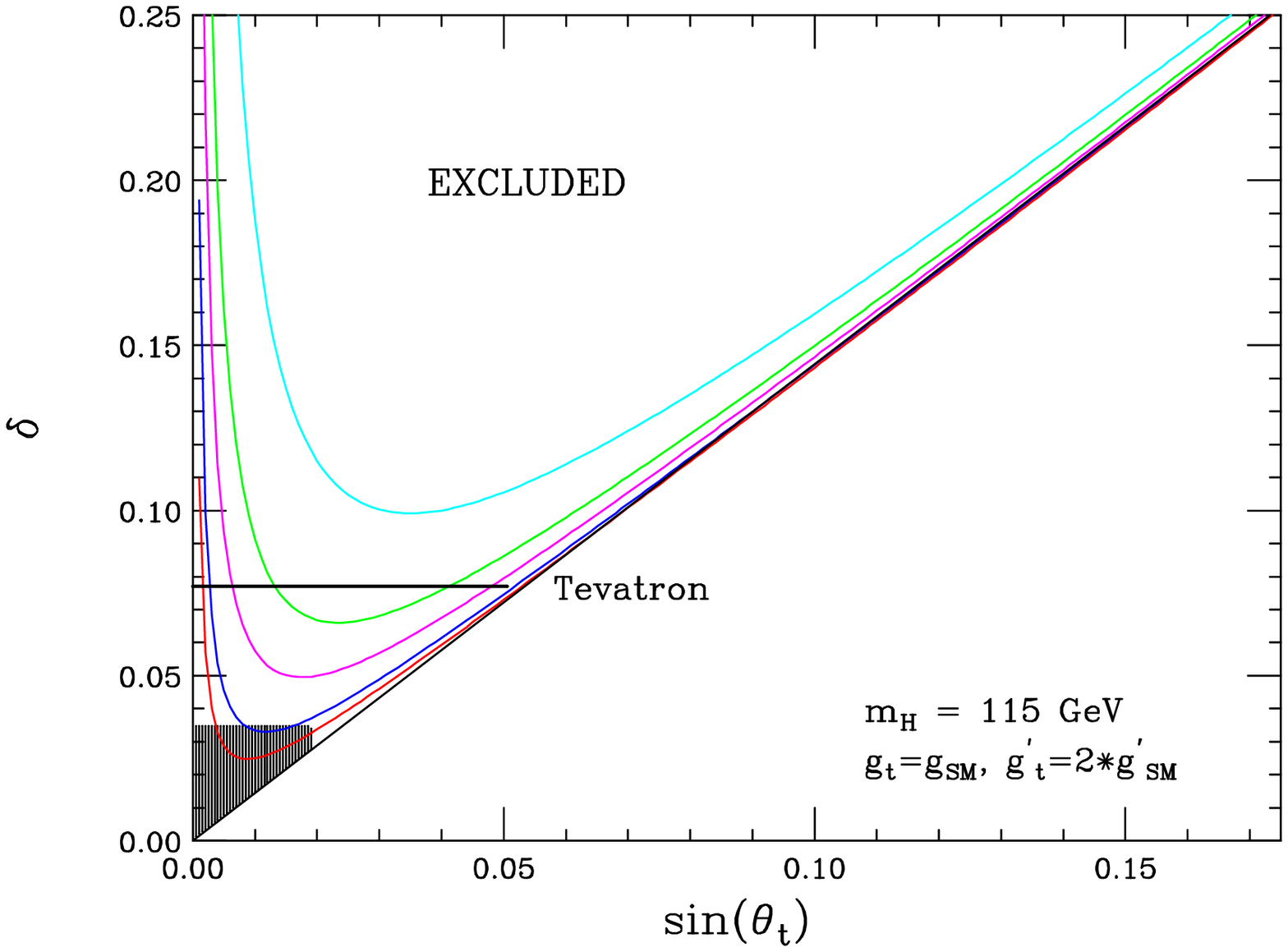,height=6.9cm,width=9.6cm,angle=0}}
\vspace*{1.0cm}
\centerline{
\psfig{figure=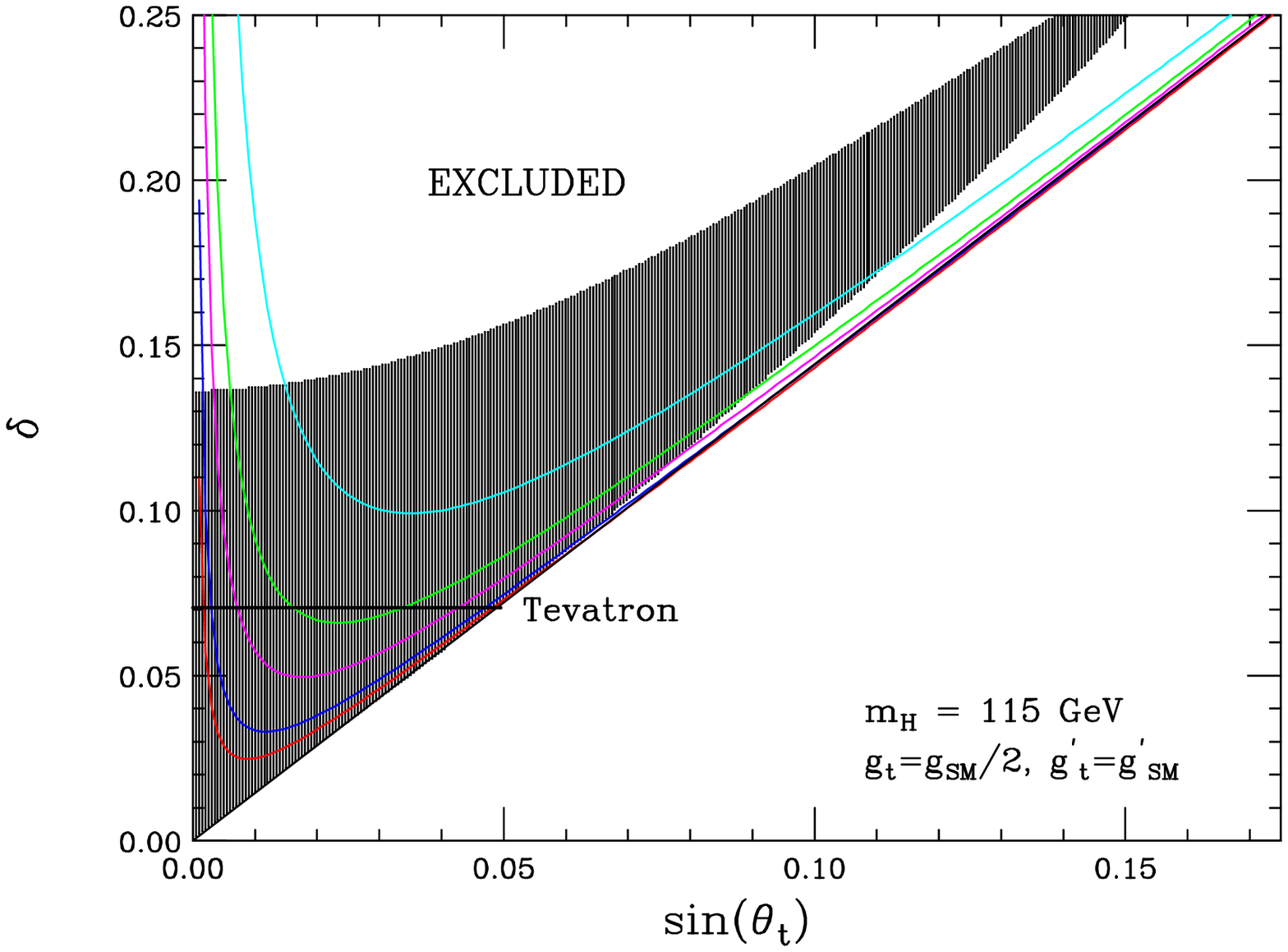,height=6.9cm,width=9.6cm,angle=0}
\psfig{figure=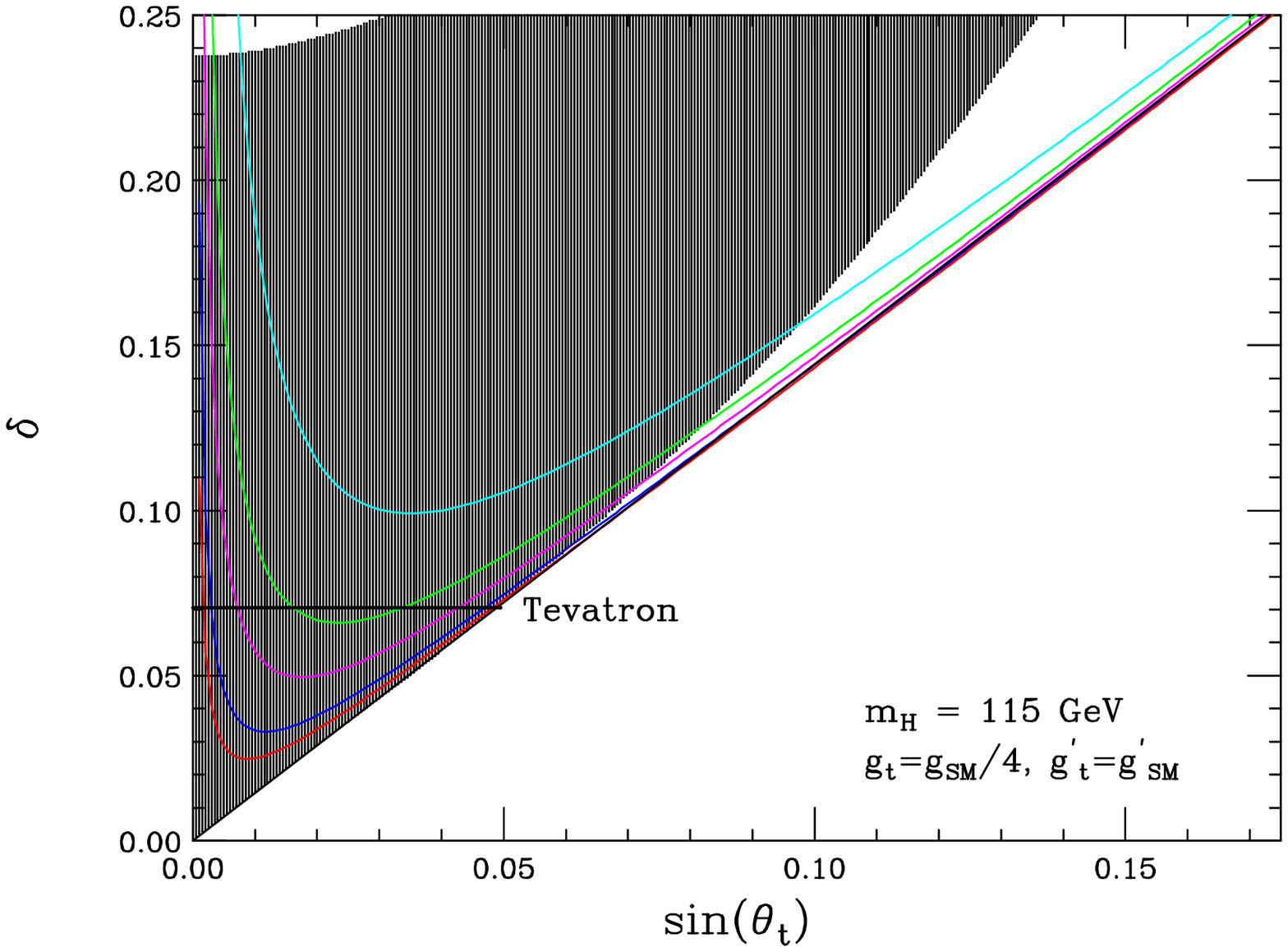,height=6.9cm,width=9.6cm,angle=0}}
\caption{Fit to the EW precision data varying ${\rm sin}
\left(\theta_t\right)$ and $\delta$, for various choices of $g_t$ and 
$g_{t^{'}}$ as indicated.  The diagonal line represents the bound 
$\delta/\sqrt{2} \leq {\rm tan}\left(\theta_t\right)$, the horizontal line denotes 
the 95\% CL bound from $B_h$ production at the Tevatron, and the series of 
curved lines corresponds to the $t^{'}$ mass $m_{t^{'}}$ as a function of 
$\delta$ and ${\rm sin}\left(\theta_t\right)$; from top to bottom, they 
represent $m_{t^{'}} = 5$, 7.5, 10, 15, and 20 TeV.  The shaded regions are 
allowed by the EW fit.}
\label{coupvar}
\end{figure}

\begin{figure}[htbp]
\centerline{
\psfig{figure=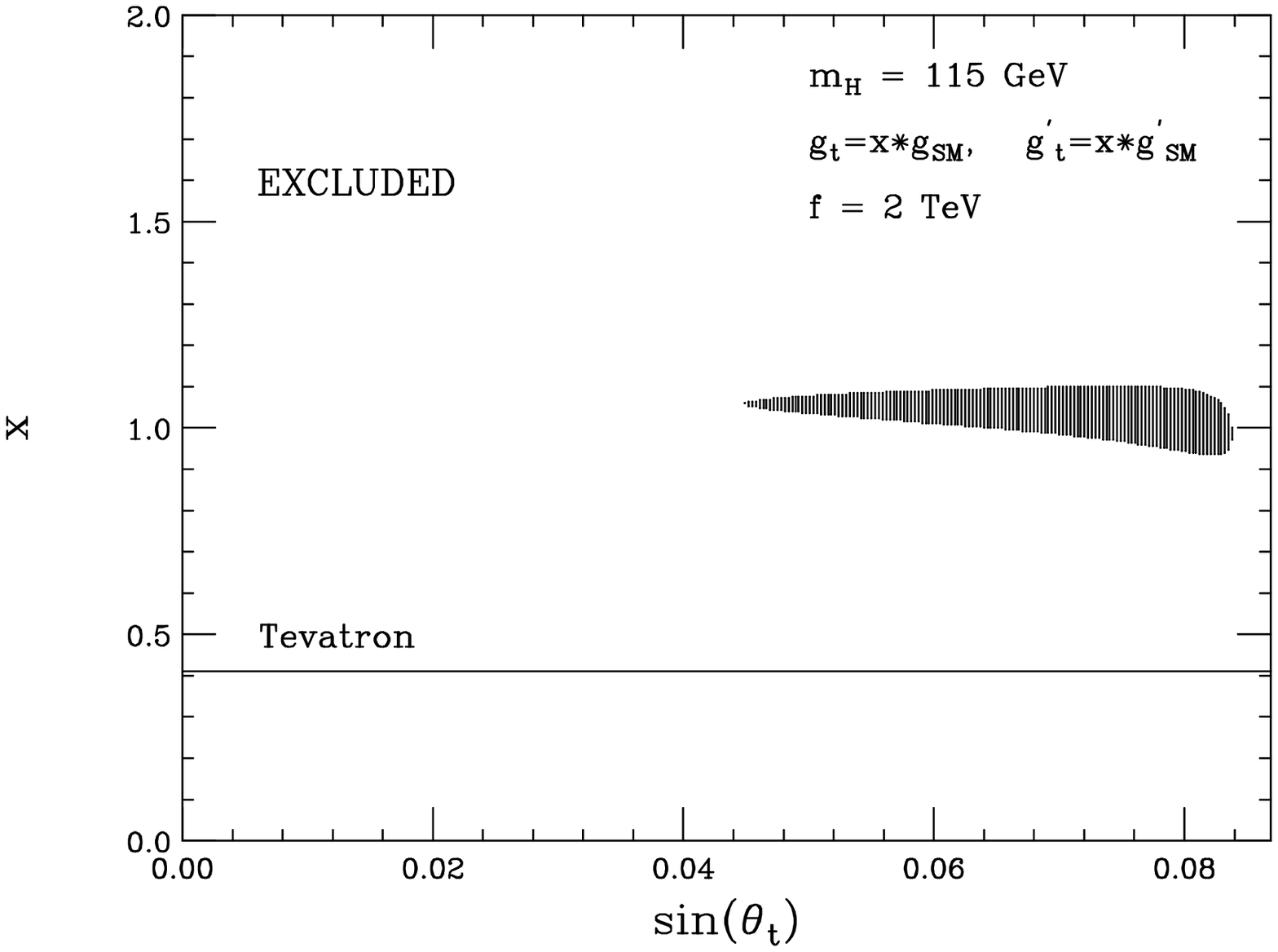,height=6.9cm,width=9.6cm,angle=0}
\psfig{figure=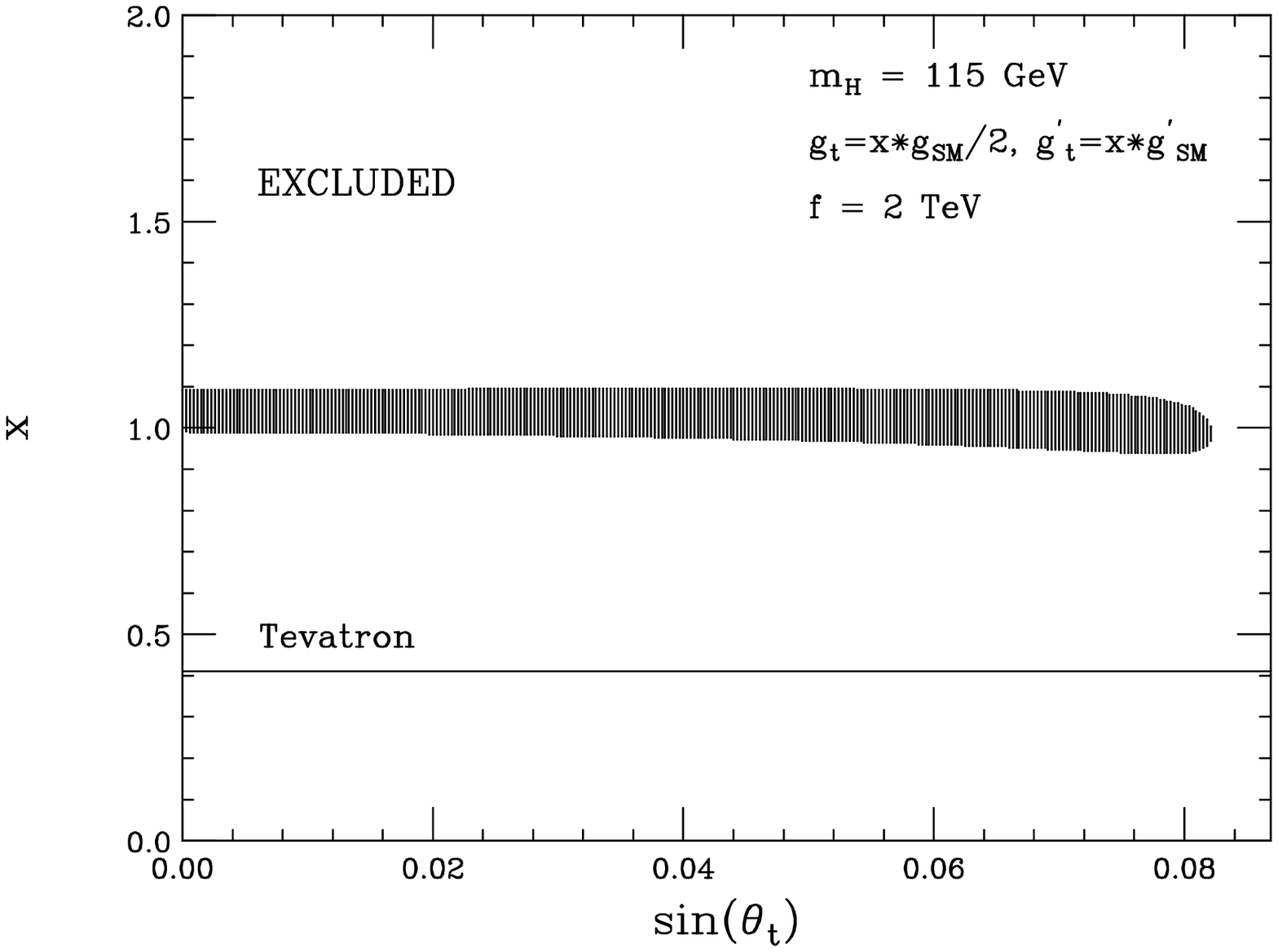,height=6.9cm,width=9.6cm,angle=0}}
\vspace*{1.0cm}
\centerline{
\psfig{figure=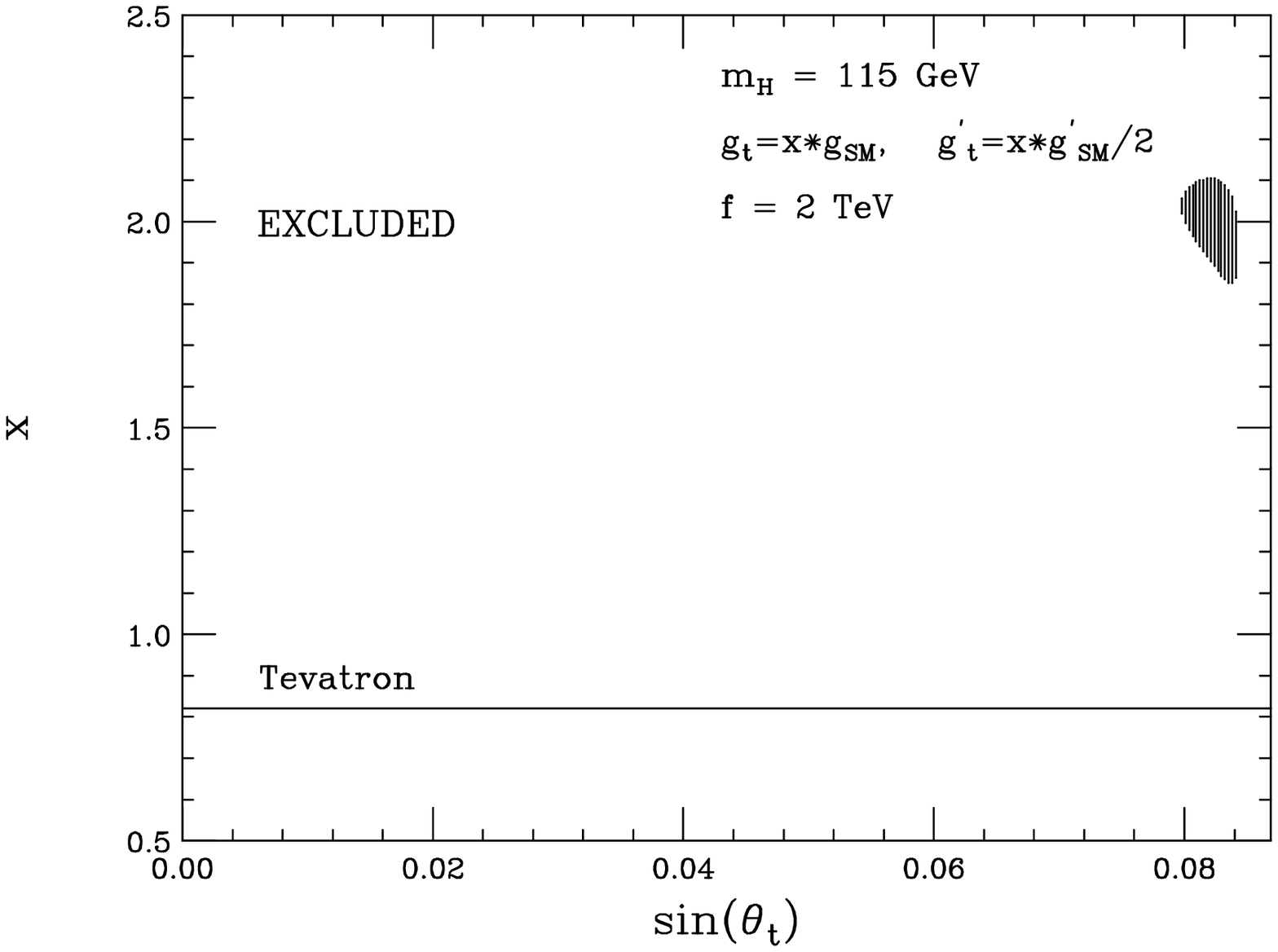,height=6.9cm,width=9.6cm,angle=0}}
\caption{Fit to the EW precision data varying both ${\rm sin}
\left(\theta_t\right)$ and $x$, for $f=2$ TeV.  The shaded regions are allowed 
by the EW fit, while the areas above the horizontal lines are forbidden 
by Tevatron direct search constraints.  The rightmost boundary of the
figure is 
determined by the constraint $\delta/\sqrt{2} \leq {\rm tan}\left(\theta_t\right)$.}
\label{xf2plots}
\end{figure}

We now study the case where the couplings $g_t$ and $g^{'}_t$
are varied within natural ranges away 
from their corresponding SM values.
For this and all later discussions we assume 
$m_H=115$ GeV.  Shown in Fig.~\ref{coupvar} are the results 
of the EW fits for the following four coupling choices: (1) $g_t=g$ and 
$g^{'}_{t}=g^{'}/2$; (2) $g_t=g$ and $g^{'}_{t}=2g^{'}$; (3) 
$g_t = g/2$ and $g^{'}_{t} = g^{'}$; (4) $g_t = g/4$ and $g^{'}_{t} = g^{'}$.  
We see from the top two figures that the EW constraints are strongest 
in the case where the Tevatron constraints are weakened.  As can be seen 
from Fig. 2, the Tevatron limits are the most stringent 
when $g^{'}_{t}/g^{'} \approx 1$, and decrease
away from this region.  This region is exactly where 
$\delta\rho_G$ vanishes, as can be seen from Eq.~\ref{drhodef}; away 
from this point the contribution of $\delta\rho$ to the EW fit becomes 
more important.  For case (1),  the EW 
precision constraints require $f \gsim 3.5$ TeV, while in case (2) this 
bound is strengthened to $f \gsim 7.2$ TeV.  These correspond to limits 
of $m_{t^{'}} > 7$ TeV and $m_{t^{'}} > 14$ TeV, respectively.  The 
strongest constraints for cases (3) and (4) arise from the 
Tevatron direct searches; the 
EW precision constraints are relatively weak for these parameter choices.  
To further demonstrate that bounds from EW data become strong in precisely 
those regions in which the Tevatron constraints decrease, we present in 
Fig.~\ref{xf2plots} the results of fits where we set $f=2$ TeV and vary both 
${\rm sin}\left(\theta_t\right)$ and $x$.  The three fits we perform use the 
following definitions of $x$: (1) $g_t = xg$ and $g^{'}_t = xg^{'}$; 
(2) $g_t = xg/2$ and $g^{'}_t = xg^{'}$; (3) $g_t = xg$ and $g^{'}_t 
= xg^{'}/2$.  These variations in the couplings within natural ranges
provide for a thorough exploration of the model parameter space.
In all three cases the parameter values allowed by the EW fit 
are clustered very near $g^{'}_{t} = g^{'}$, and are excluded by the 
Tevatron constraints.  We obtain identical results if we choose $f=3$ TeV.
Combining the results of these fits, we can conclude that $f \gsim 3.5$ TeV 
throughout the full parameter space of the model. 
This implies a {\it minimum} mass 
of $m_{t^{'}} = 7$ TeV for the $t^{'}$ state, which is beyond the
reach of the LHC.

\vspace*{-0.55cm}
\noindent

\vspace*{-0.55cm}
\noindent

\begin{figure}[htbp]
\centerline{
\psfig{figure=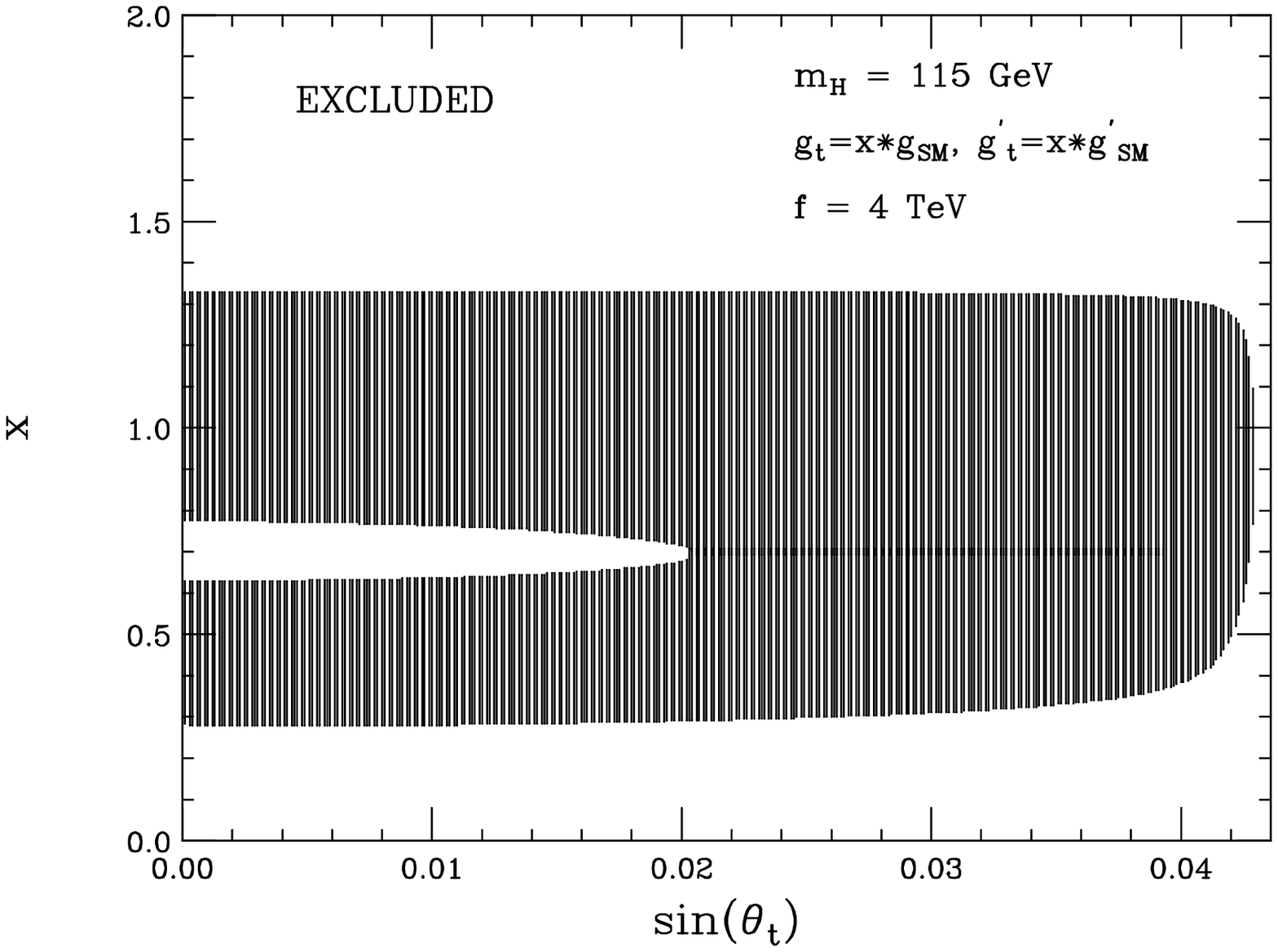,height=6.9cm,width=9.6cm,angle=0}
\psfig{figure=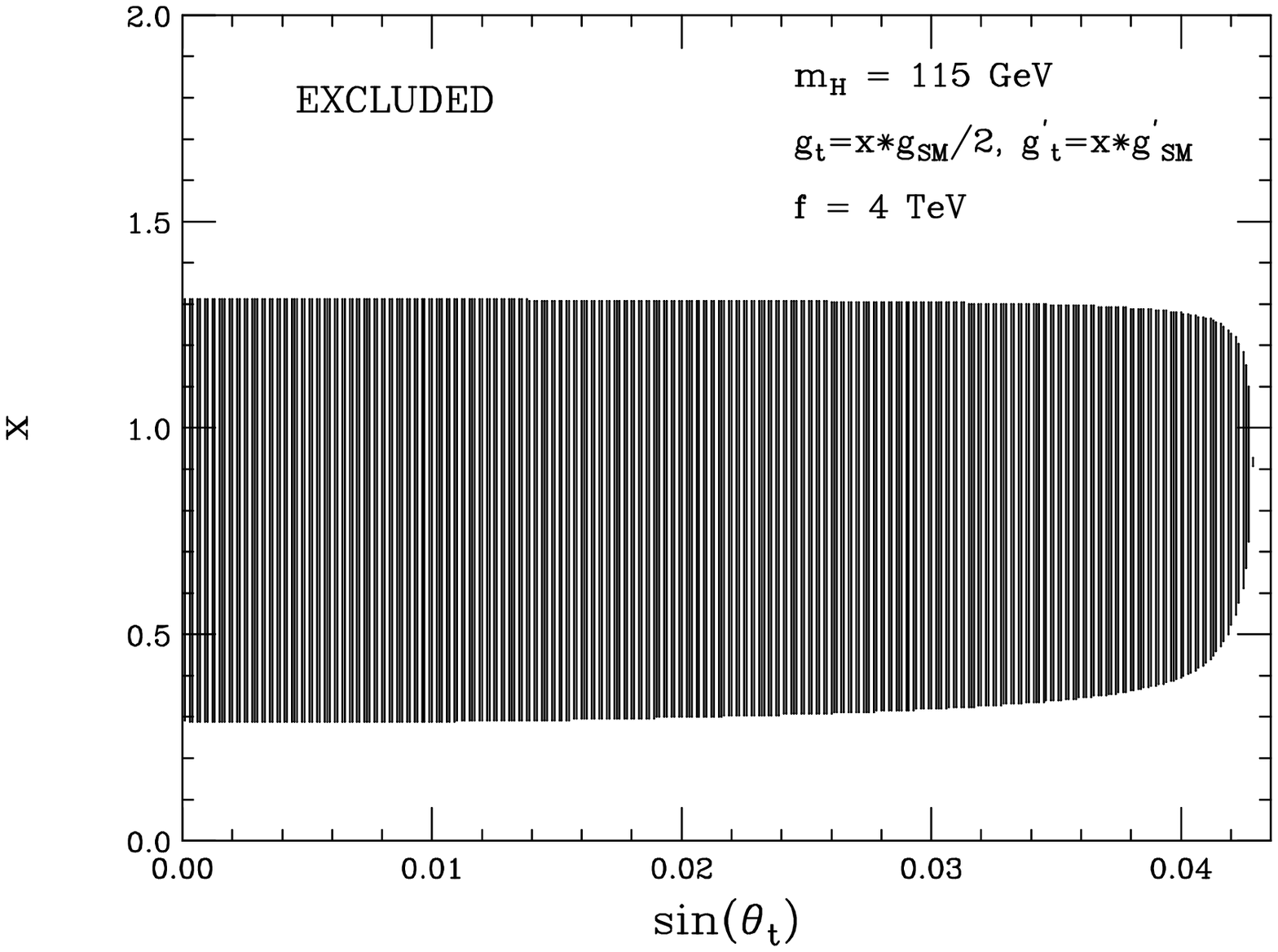,height=6.9cm,width=9.6cm,angle=0}}
\vspace*{1.0cm}
\centerline{
\psfig{figure=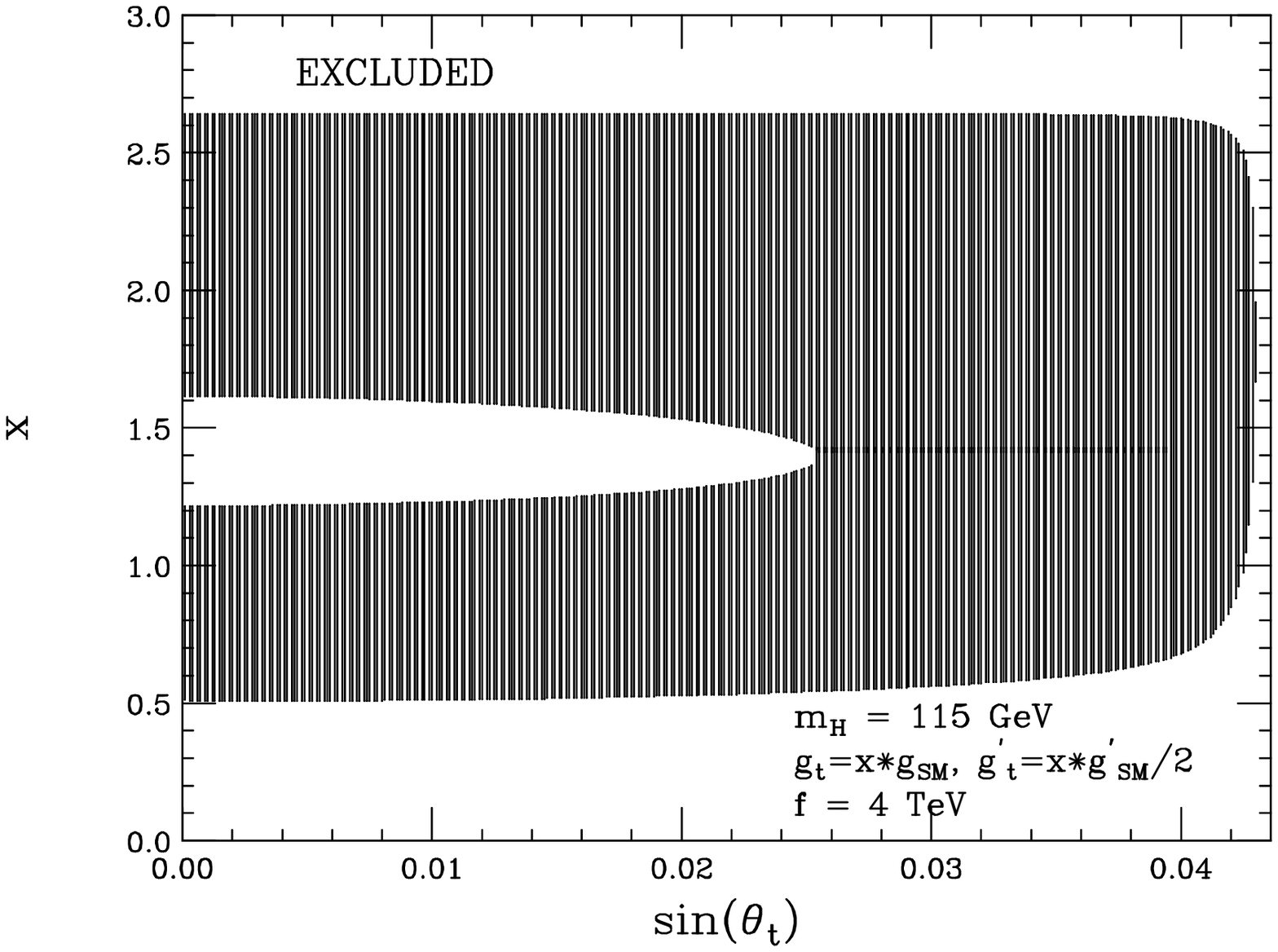,height=6.9cm,width=9.6cm,angle=0}}
\caption{Fit to the EW precision data varying both ${\rm sin}
\left(\theta_t\right)$ and $x$, for $f=4$ TeV.  The shaded regions are allowed 
by the EW fit.  The rightmost boundary of the figure is 
determined by the constraint $\delta/\sqrt{2} \leq {\rm tan}\left(\theta_t\right)$.}
\label{xf4plots}
\end{figure}

\begin{figure}[htbp]
\centerline{
\psfig{figure=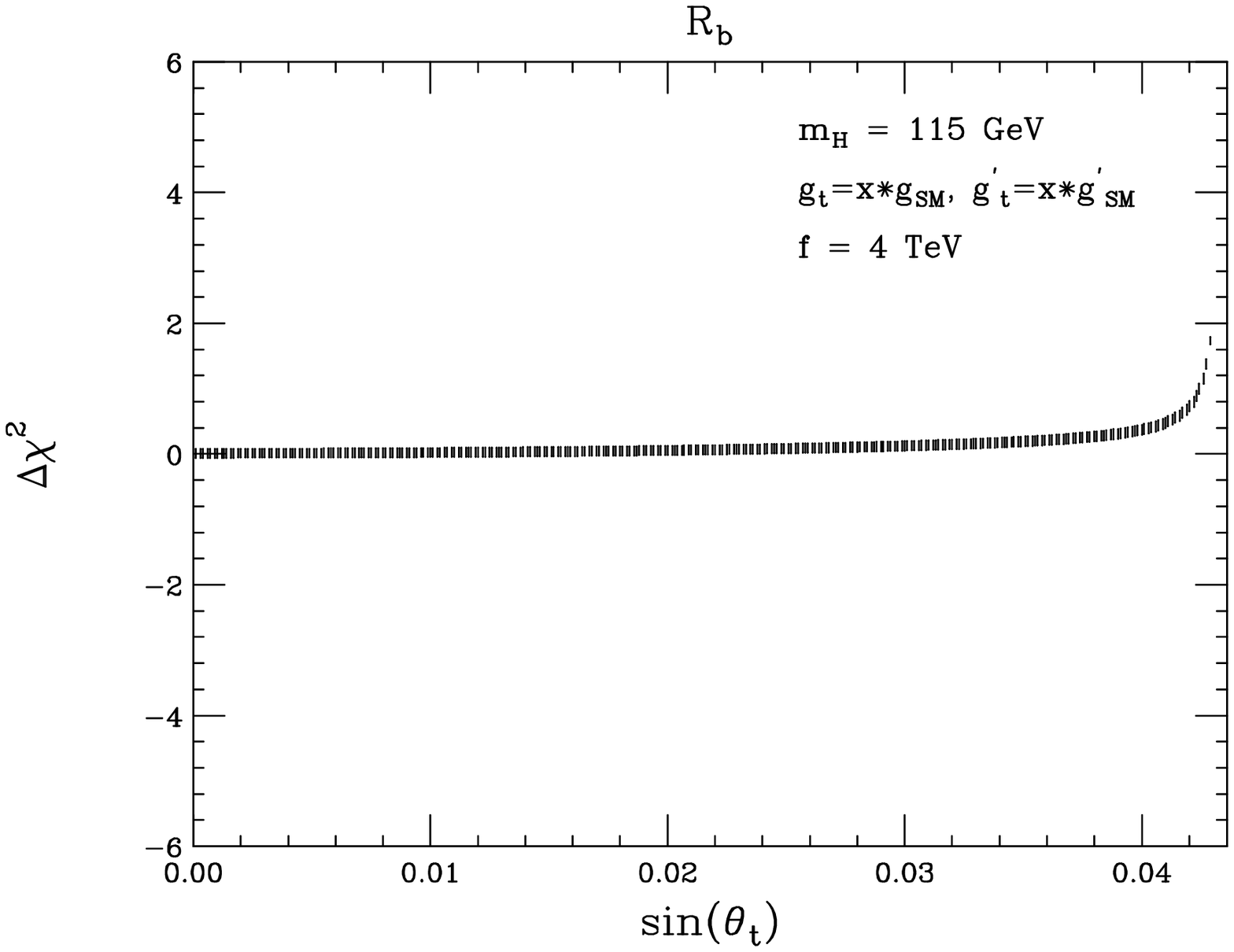,height=6.5cm,width=8.8cm,angle=0}
\psfig{figure=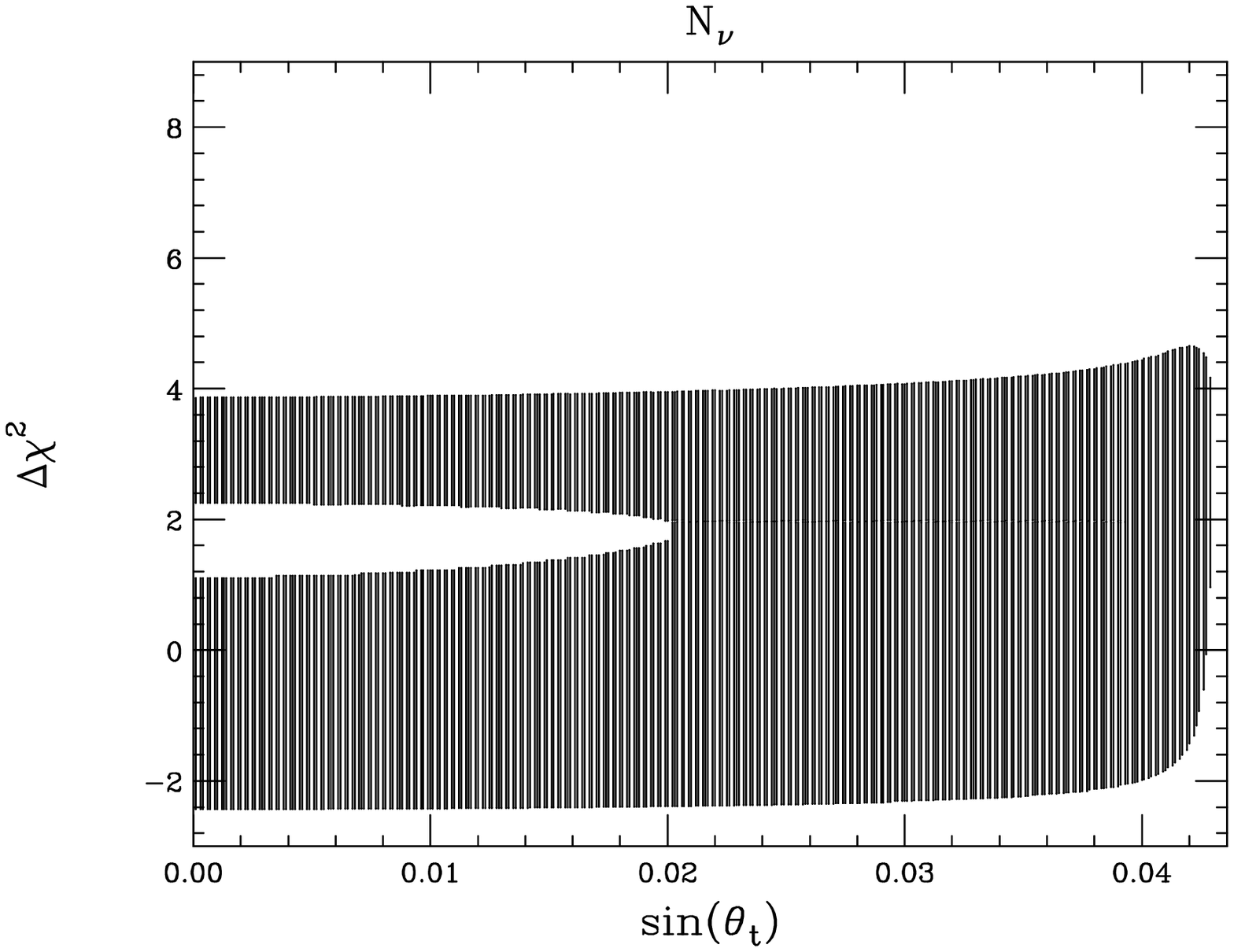,height=6.5cm,width=8.8cm,angle=0}}
\vspace*{0.5cm}
\centerline{
\psfig{figure=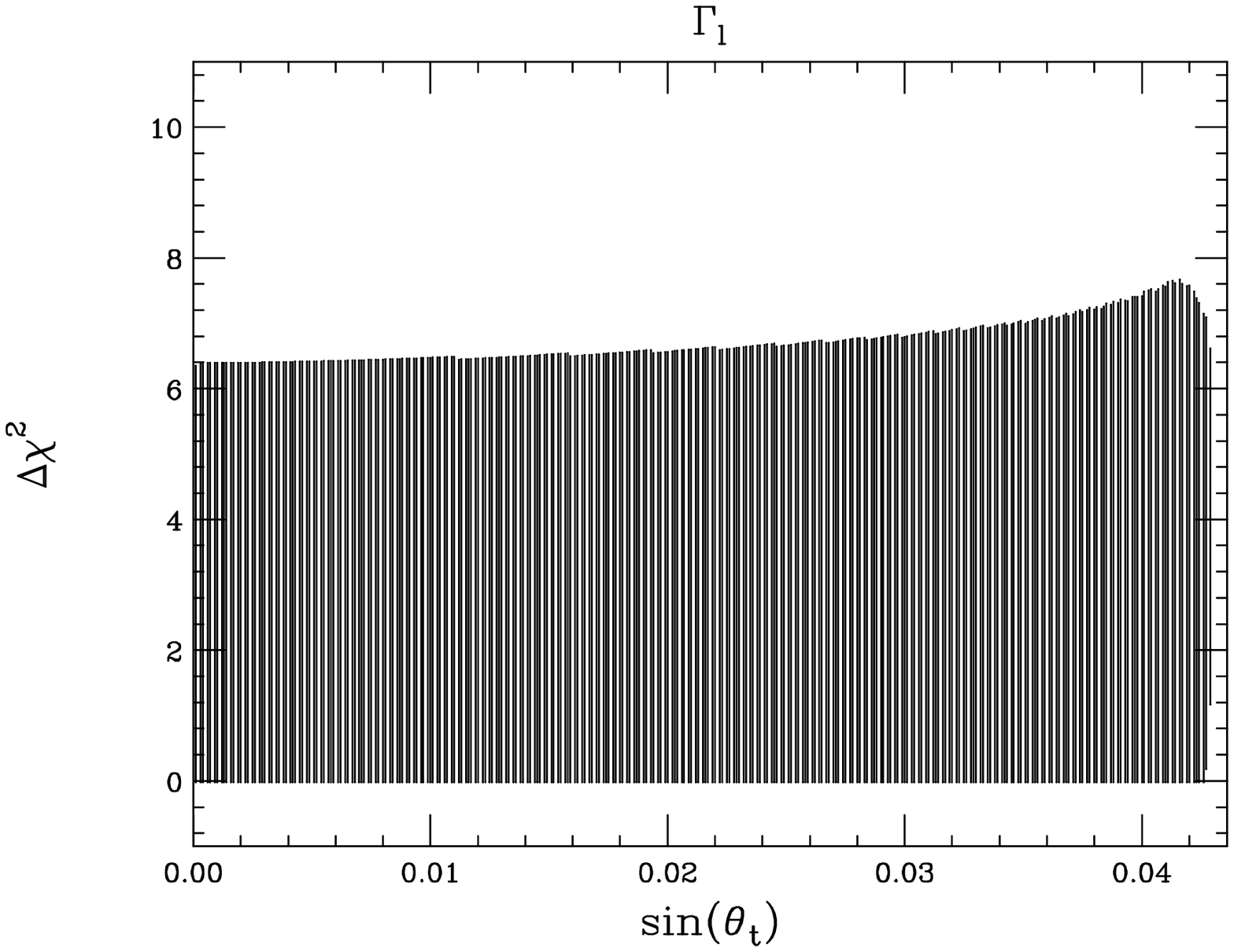,height=6.5cm,width=8.8cm,angle=0}
\psfig{figure=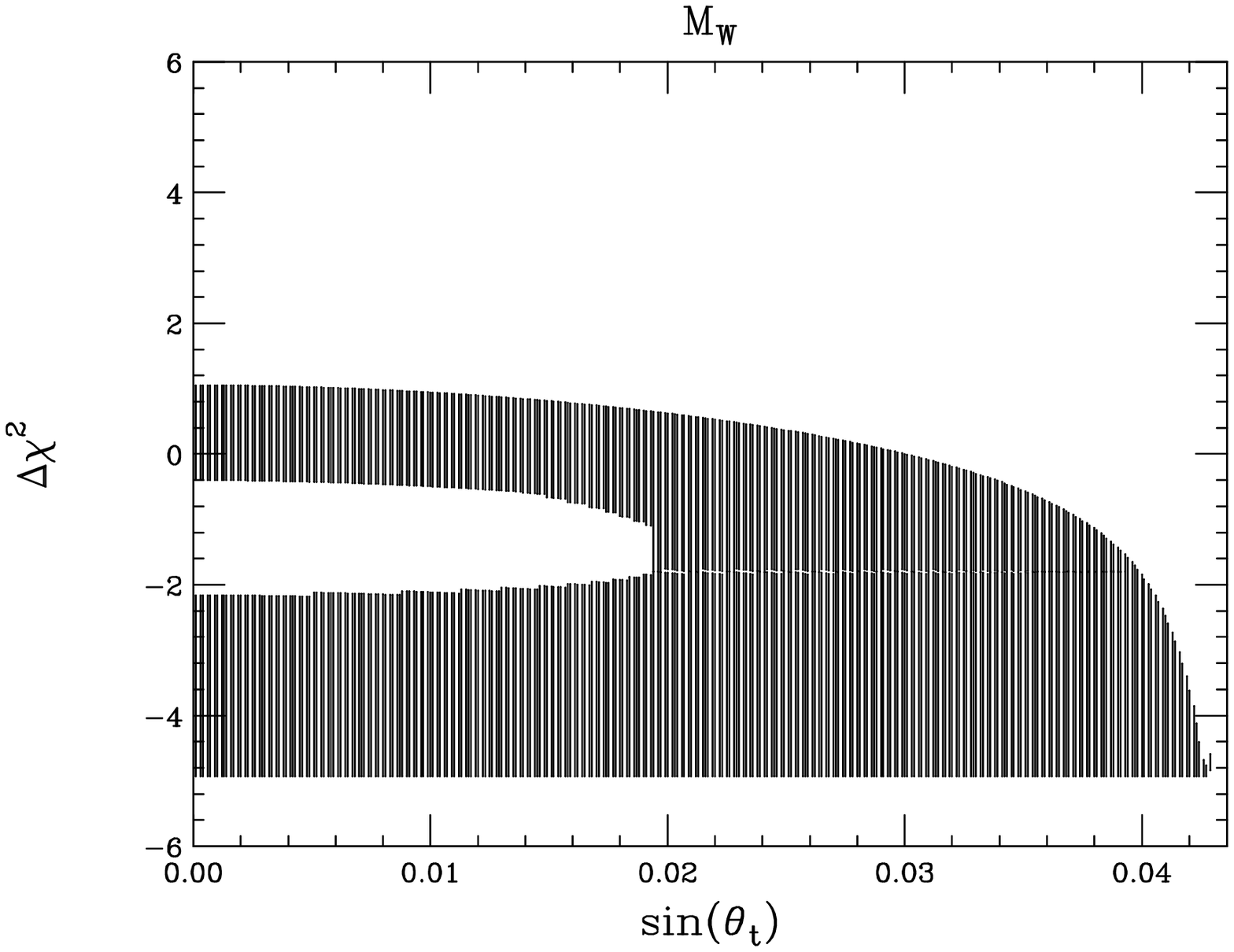,height=6.5cm,width=8.8cm,angle=0}}
\vspace*{0.5cm}
\centerline{
\psfig{figure=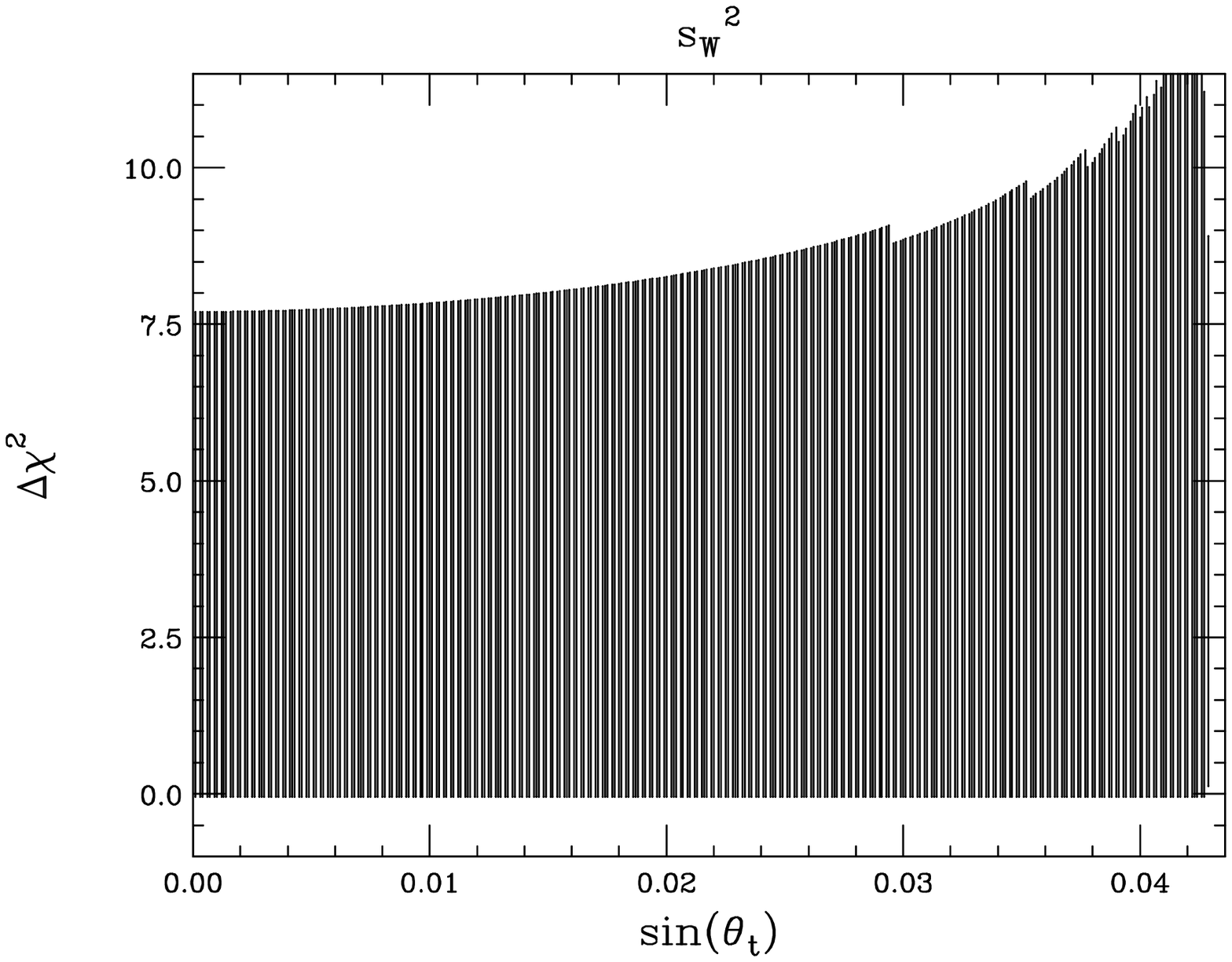,height=6.5cm,width=8.8cm,angle=0}}
\caption{The $\Delta\chi^2$ values for several observables as functions of 
${\rm sin}\left(\theta_t\right)$ for the case 
$g_t = xg$ and $g^{'}_t = xg^{'}$; 
included are all $x$ values that satisfy the precision EW constraints
as shown in the previous figure.}
\label{xf4chi}
\end{figure}

Having established that $f \lsim 3.5$ is excluded by the data, 
we now set $f=4$ TeV, which is allowed 
by both EW precision constraints and Tevatron direct search limits.  We study 
the fit to the EW precision data as we vary both 
${\rm sin}\left(\theta_t\right)$ and $x$; we choose the same three cases 
for the relations of the couplings
as examined above.  The results of these fits are presented in 
Fig.~\ref{xf4plots}.  We see that a far larger region of the coupling 
space is consistent with the data 
in this case, and that the EW data allows significant deviations 
from the $g^{'}_{t} = g^{'}$ limit.  To examine the contributions of the 
various observables to the fit, we present in Fig.~\ref{xf4chi} the 
$\Delta\chi^2$ values for the various observables obtained in the $g_t = xg$, 
$g^{'}_t = xg^{'}$ fit;
$\Delta\chi^2$ is the difference between the $\chi^2$ values obtained in the 
Littlest Higgs model and in the SM.  Included in these figures are all values 
of $x$ which satisfy the bounds imposed by the EW data.  We have not 
presented our results for 
the observables $R_c$, $A_c$, and $A_b$ because their 
$\Delta\chi^2$ values are very nearly zero throughout the entire parameter 
space; \ie, they have no influence on the fit.  All of the $\Delta\chi^2$ 
ranges exhibit a dip either upwards or downwards at the 
right edge of these figures.  
These arise from the $t$ and $t^{'}$ contributions to either 
$\delta\rho$ or, for $R_b$, the $Z\bar{b}b$ vertex.  Although $m_{t^{'}}$ 
becomes large as ${\rm sin}\left(\theta_t\right) \rightarrow \delta/\sqrt{2}$, 
since the $t^{'}$ mixes with the top quark it does not decouple, in the 
sense discussed above, for some 
parameter space regions. We see that $M_{W}$ is generally better predicted by 
the Littlest Higgs model than the SM; this remains true for other parameter 
choices.  The shift in $M_{W}$ depends only on the coupling of the 
fermions to the $SU(2)_1$ gauge bosons, and not on their coupling to the 
hypercharge bosons; this will therefore remain true for other 
choices of fermion transformation assignments.  We also see that while 
$R_b$ is not 
affected until ${\rm sin}\left(\theta_t\right) \rightarrow \delta/\sqrt{2}$, 
both $\Gamma_l$ and $s_{W,eff}^{2,lep}$ are very sensitive to the deviations 
predicted by the Littlest Higgs, and that the agreement between the 
predicted values for these observables and the experimental measurements 
is always worse than in the SM.

\vspace*{-0.55cm}
\noindent

\vspace*{-0.55cm}
\noindent

\begin{figure}[htbp]
\centerline{\psfig{figure=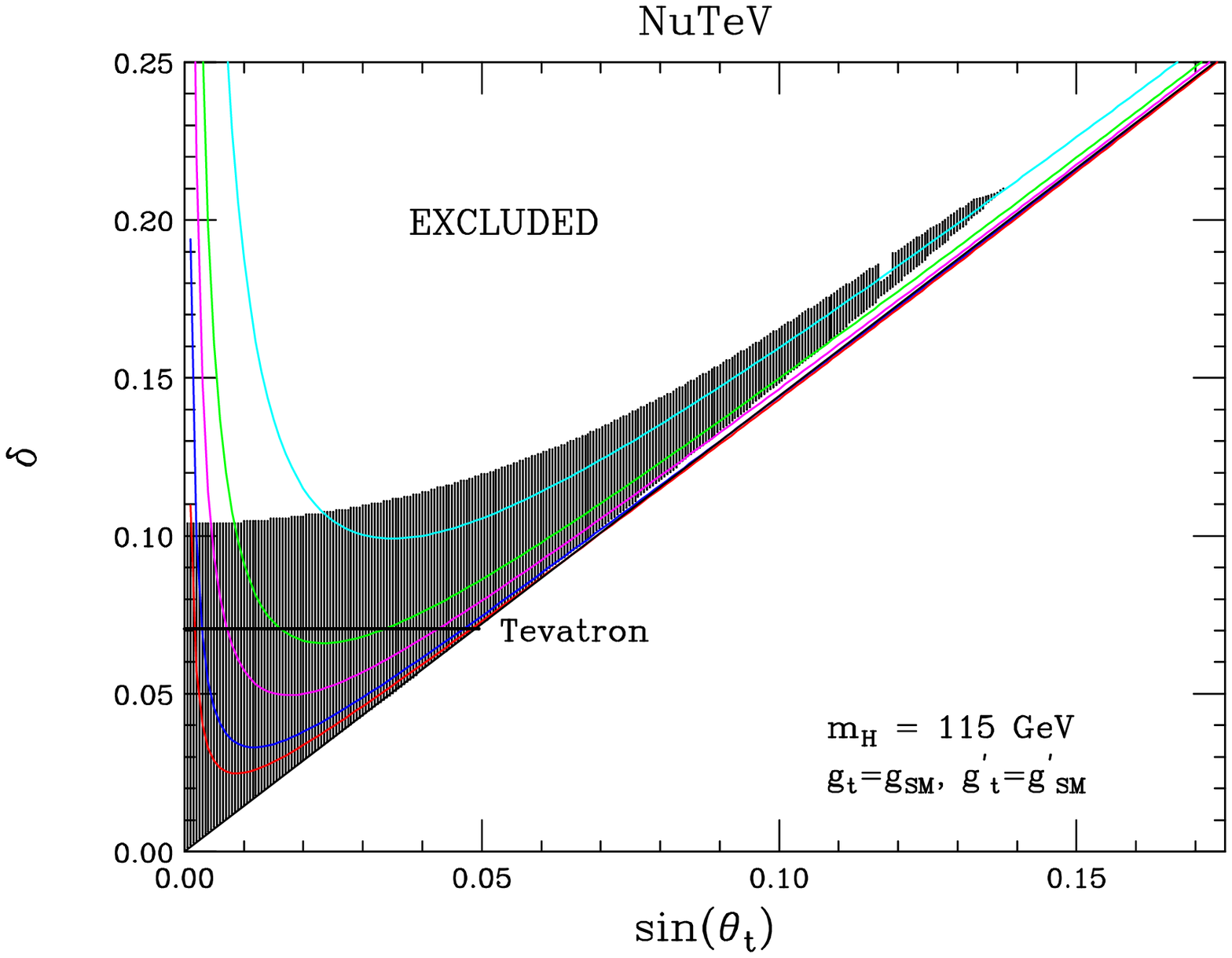,height=8.5cm,width=11.5cm,angle=0}}
\vspace*{1.0cm}
\centerline{\psfig{figure=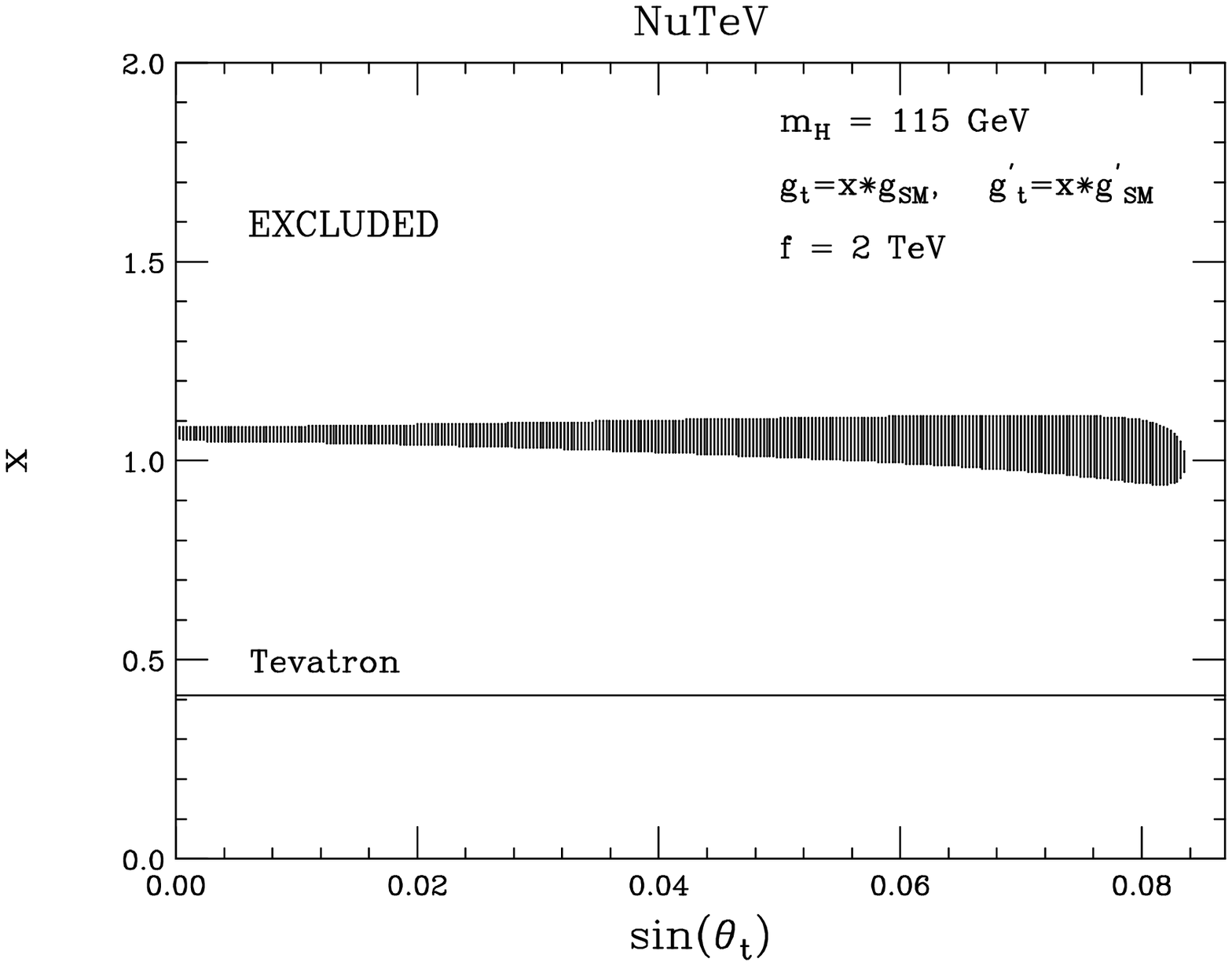,height=8.5cm,width=11.5cm,angle=0}}
\caption{Fit to the EW precision data including the NuTeV measurement of the 
on-shell weak mixing angle.  The top figure studies the variation with 
$\delta$ and ${\rm sin}\left(\theta_t\right)$, while the lower figure 
shows the 
dependence on ${\rm sin}\left(\theta_t\right)$ and $x$ with $f=2$ TeV.}
\label{nutev1}
\end{figure}

Finally, we discuss the effect on the fit of the NuTeV measurement of the 
on-shell value of the weak mixing angle $s_{W,N}^{2,os}$.  The NuTeV 
result currently disagrees with that derived from $Z$-pole data by 
approximately 3 standard deviations~\cite{Zeller:2001hh}.  Although this is 
possibly a signal of new physics, several more mundane explanations, such 
as larger parton distribution function uncertainties than those assumed by 
the NuTeV collaboration, asymmetric strange and charm sea quark 
distributions, or nuclear effects, have 
been suggested~\cite{Nudisc}.  With this caveat in mind, we will assume 
both the central value and error determined by the NuTeV collaboration, 
and repeat our analysis including the NuTeV result.  
To demonstrate that the 
allowed range of $f$ is unaffected by the inclusion of the NuTeV 
measurement, we display the results of re-performing 
the following two analyses in 
Fig.~\ref{nutev1}: (1) varying $\delta$ and ${\rm sin}\left(\theta_t\right)$ 
while setting $g_t=g$ and $g^{'}_{t} = g^{'}$, the analog of the fit 
presented in Fig.~\ref{higgsvar}; (2) varying ${\rm sin}\left(\theta_t\right)$ 
and $x$ with $f=2$ TeV, and $x$ defined by $g_t=xg$ and $g^{'}_{t} = xg^{'}$, 
the analog of the result shown in Fig.~\ref{xf2plots}.  Comparing these 
figures with those presented previously, we see that the constraint on 
$f$ is unchanged by the addition of the NuTeV data.  Although the full EW 
data set allows a slightly larger range of parameters here than before, the 
Tevatron constraint again excludes these regions.  For other
coupling choices, the NuTeV measurement again either very slightly increases 
or decreases the size of the allowed parameter space; no large swath of new 
parameter values are allowed.

\vspace*{-0.8cm}
\noindent

\begin{figure}[htbp]
\centerline{\psfig{figure=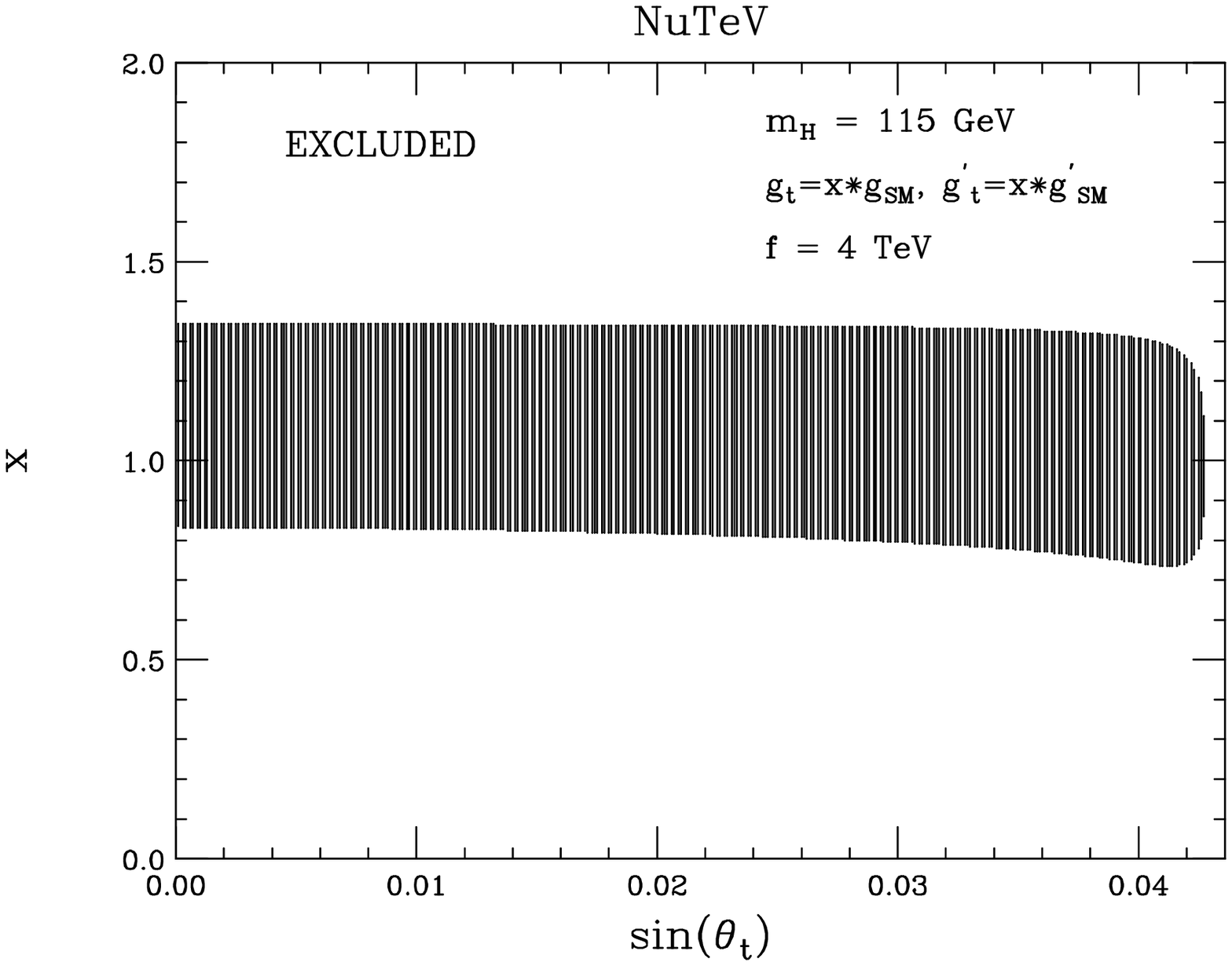,height=8.5cm,width=11.5cm,angle=0}}
\vspace*{1.0cm}
\centerline{\psfig{figure=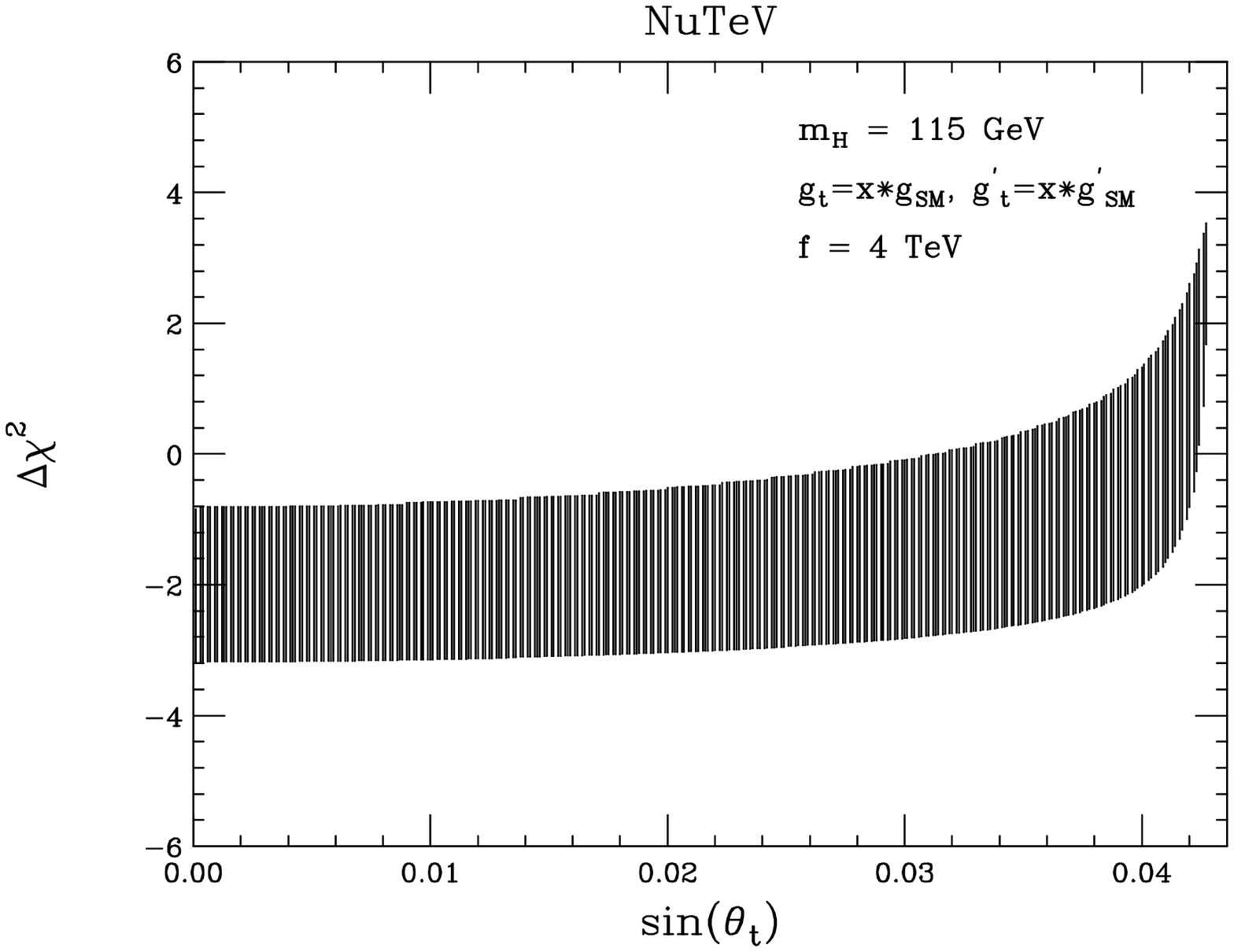,height=8.5cm,width=11.5cm,angle=0}}
\caption{Fit to the EW precision data including the NuTeV measurement of the 
on-shell weak mixing angle; we have set $f=4$ TeV and $g_t=xg$, 
$g^{'}_{t} = xg^{'}$.  The top figure shows the fit to the EW data, while 
the lower figure illustrates the $\chi^2$ shift from the SM for those 
values of $x$ and ${\rm sin}\left(\theta_t\right)$ which satisfy the 
EW constraints.}
\label{nutev2}
\end{figure}

We next set $f=4$ TeV, $g_t=xg$ and $g^{'}_{t} = xg^{'}$, and vary 
both ${\rm sin}\left(\theta_t\right)$ and $x$; the results are presented 
in Fig.~\ref{nutev2}.  We display the allowed region in the 
$x-\sin(\theta_t)$ plane as well as the values of $\Delta\chi^2$ 
for $s_{W,N}^{2,os}$.  
The Littlest Higgs model shifts $s_{W,N}^{2,os}$ in the direction preferred 
by the NuTeV measurement throughout most of the entire parameter space; 
this remains true for other choices of the couplings.  The size of
the allowed region is smaller than in the case where the NuTeV data
is not included.  The shifts induced by the Littlest Higgs model ($
\approx 0.5-1.5$ $\sigma$)
cannot compensate for the large $\approx 3$ $\sigma$ deviation
of the SM prediction from the NuTeV results.

\vspace*{-0.8cm}
\noindent

Before concluding we note that we have not discussed the possible influence 
of measurements of Atomic Parity Violation (APV) on the fits. Until recently 
the experimental value of the weak charge, $Q_W$, differed from the 
predictions of the SM by about $2\sigma$ {\cite {APV1}}. After a series of 
detailed theoretical calculations including various atomic effects 
it now seems that experiment and theoretical predictions are in 
agreement {\cite {APV2}}.  However, since the 
theoretical calculation of $Q_W$ may still be in flux we do not include 
this observable in our fits.  After computing the corrections to
$Q_W$ in this model and scanning the allowed 
regions from our previous fits, we find that either sign for
the shift of $Q_W$ can be generated.

\section{Conclusions}

Little Higgs models are a novel attempt to address the hierarchy
problem. These theories predict the existence of a host of new 
particles at the TeV scale, which are necessary to cancel the 
quadratic divergences of the Standard Model.  Here, we have examined
the simplest version of 
these models, the Littlest Higgs, which contains 
four new gauge bosons, a weak isosinglet quark, $t'$,  with $Q=2/3$, 
as well as an isotriplet scalar field with their masses being constrained 
by naturalness requirements. In this paper, we have considered the 
contributions of these new states to precision electroweak 
observables, and have examined the additional constraints provided by the
direct searches for new particles 
at the Tevatron.  We have performed a thorough exploration of the 
parameter space of this model and have found that there are 
small regions allowed by the precision data where the parameters take 
on their natural values.  When the additional limits provided by the 
Tevaton are included, these regions are no longer permitted. 
By combining the direct and indirect effects of these new states 
we have constrained  the `decay constant'$f$ to rather large values 
$\gsim 3.5$ TeV; similarly, to satisfy both sets of data  
$m_{t'}\gsim 7 $ TeV which is far beyond the reach of direct searches at 
the LHC. These bounds imply that significant fine-tuning must be present in 
order for this model to resolve the hierarchy. We thus find that the Littlest 
Higgs model is tightly constrained by the combination 
of precision electroweak data and direct Tevatron searches.

\bigskip
\noindent
{\Large\bf Acknowledgements}

The authors would like to thank Nima Arkani-Hamed, Gustavo Burdman, 
Alex Kagan, Ann Nelson, Michael Peskin, Aaron Pierce and Jay Wacker 
for useful discussions related to this work.

\noindent
{\it Note Added}:  A related work \cite{csaki}, which studied the effects of
mixing in the gauge sector in the Littlest Higgs model, appeared while this
paper was being completed.

\bigskip
\noindent
{\Large\bf Appendix A}
\bigskip

\begin{figure}[htbp]
\centerline{
\psfig{figure=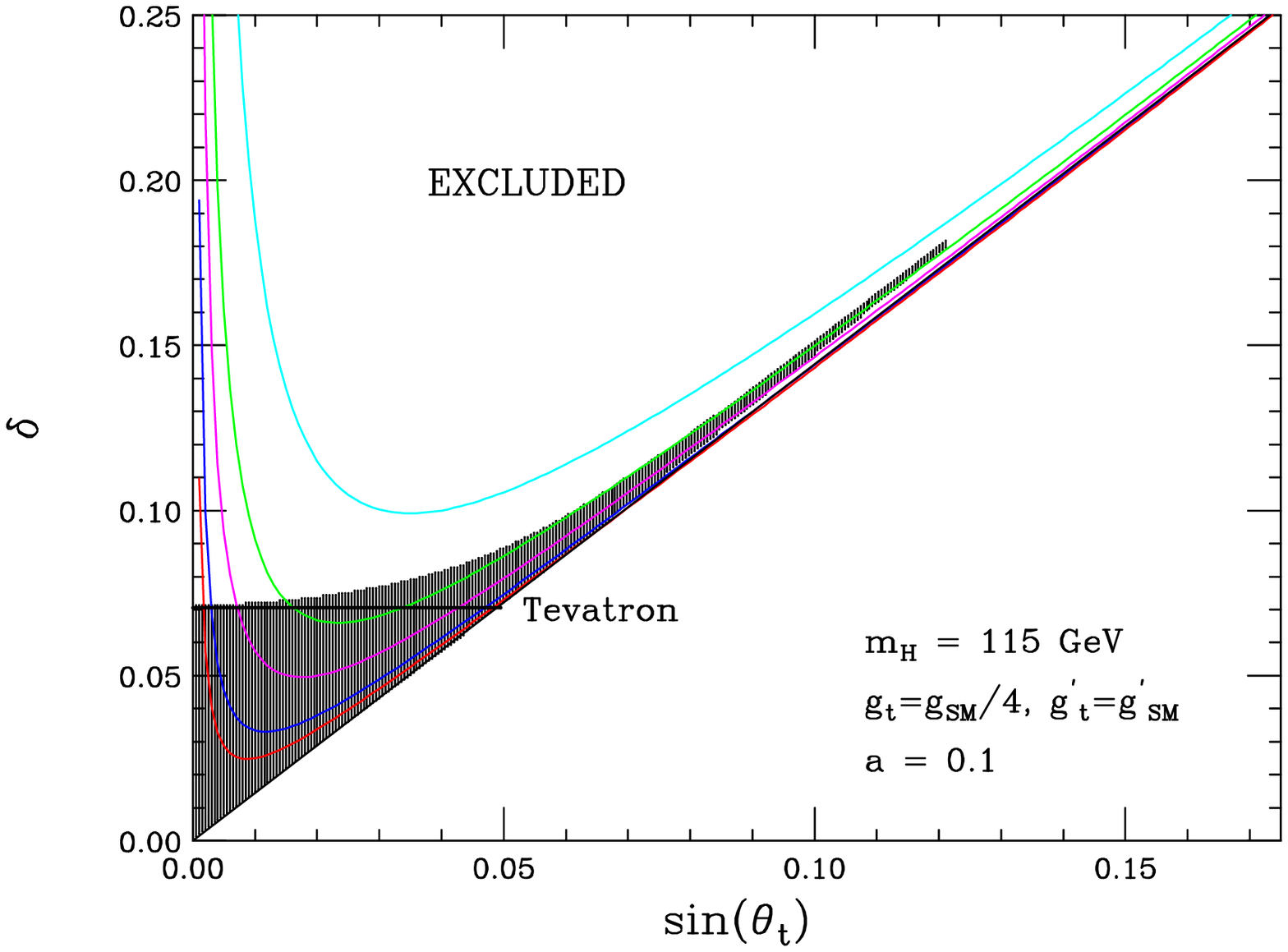,height=6.5cm,width=8.8cm,angle=0}
\psfig{figure=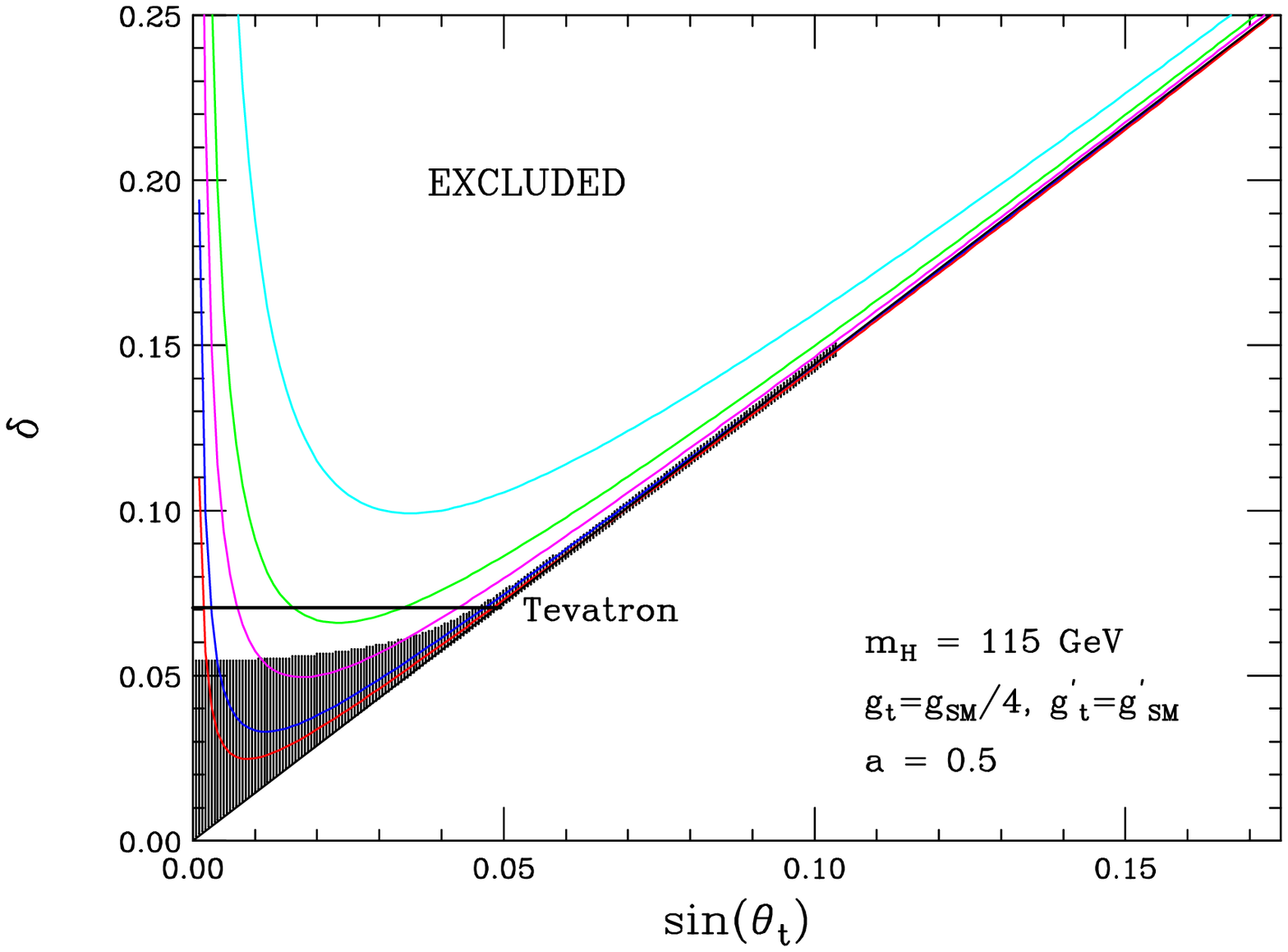,height=6.5cm,width=8.8cm,angle=0}}
\vspace*{0.5cm}
\centerline{
\psfig{figure=mh115_gt4.ps,height=6.5cm,width=8.8cm,angle=0}}
\caption{Results from the global fit to the electroweak precision
data, including the contributions from the scalar triplet vev to
the electroweak observables for various values of the parameter $a$.
Also shown are the results from the fit when the triplet vev is
not included. 
The horizontal line denotes the 95\% CL bound from $B_h$ production
at the Tevatron, and the series of curved lines corresponds to the
$t'$ mass $m_{t'}$ as a function of $\delta$ and $\sin(\theta_t)$;
from top to bottom they represent $m_{t'}=2.5,\, 5,\, 7.5,$\, and 10
TeV.  The shaded regions are allowed by the EW fit.}
\label{3vev}
\end{figure}

Here we investigate the effects of the scalar triplet vev 
on the precision electroweak observables.  The vev gives a negative 
contribution to the $\rho$-parameter which can be written as
\begin{equation}
\delta\rho_\phi = -{\delta^2\over 4} \left[ {2\lambda g_t^2\over
a(g^2+g_t^2)^2} -1 \right]^2\,,
\end{equation}
where $\lambda$ is the Higgs quartic coupling, which can be expressed in terms
of the Higgs mass via~\cite{nima}
\begin{equation}
\lambda\sim {1\over 3}\left[ {m_h\over 200~{\rm GeV}}\right]^2\,.
\end{equation}
$a$ represents a new parameter arising from the coefficient of an
ultra-violet operator; naturalness suggests that the value of $a$
lies in the range $0.1 - 1.0$.  Including this contribution
in the fit to the global electroweak data set (minus the NuTeV measurement)
yields the results displayed in Fig. \ref{3vev}.  Here, we have examined
the previously least restrictive case, as observed from Fig. 4, 
and set the values of the couplings to $g_t=g/4$, $g_t' =g'$,
while varying the parameter $a$.  For comparison, we also show our previous
results for the case when the contributions from the triplet vev are not
included.  We see that the addition of the 
contribution from the triplet vev only worsens the ability of this
model to accomodate the electroweak data set.  We have checked that the
constraints become even more restrictive as $a$ approaches unity,
and as the value of the Higgs mass increases.

\noindent

\bigskip
\noindent
{\Large\bf Appendix B}
\bigskip

We define here the loop integrals in Eq.~\ref{loopdef} which arise in 
the computation of the shift in $\delta g^{b}_L$ due to the vertex correction 
diagrams containing $t$ and $t^{'}$ states:
\begin{eqnarray}
I_1\left(m_i,m_j\right)&=& \int_{0}^{1} dx\int_{0}^{1-x} dy 
\frac{m_im_j}{\left(1-x-y\right)M_{W}^2-xyM_{Z}^2+xm_{i}^2+ym_{j}^2} \,\, , 
\nonumber \\
I_2\left(m_i,m_j\right)&=& \int_{0}^{1}dx \int_{0}^{1-x} dy \,
{\rm ln}\left[ \frac{\left(1-x-y\right)M_{W}^2-xyM_{Z}^2+xm_{i}^2
+ym_{j}^2}{\mu^2}\right] \,\, , \nonumber \\
I_3\left(m_i\right) &=& \int_{0}^{1} dx\int_{0}^{1-x} dy\, {\rm ln}\left[
\frac{\left(1-y\right)M_{W}^2 -x\left(1-x-y\right)M_{Z}^2+ym_{i}^2}{\mu^2} 
\right]
\,\, , \nonumber \\
I_4\left(m_i\right) &=& \int_{0}^{1} dx \, x \, {\rm ln}\left[ 
\frac{xM_{W}^2+\left(1-x\right)m_{i}^2}{\mu^2} \right] \,\, ,
\label{intsdef}
\end{eqnarray}
where $\mu$ is an arbitrary renormalization scale that cancels from the
final expression after all the individual diagrams are summed over.

%
\def\MPL #1 #2 #3 {Mod. Phys. Lett. {\bf#1},\ #2 (#3)}
\def\NPB #1 #2 #3 {Nucl. Phys. {\bf#1},\ #2 (#3)}
\def\PLB #1 #2 #3 {Phys. Lett. {\bf#1},\ #2 (#3)}
\def\PR #1 #2 #3 {Phys. Rep. {\bf#1},\ #2 (#3)}
\def\PRD #1 #2 #3 {Phys. Rev. {\bf#1},\ #2 (#3)}
\def\PRL #1 #2 #3 {Phys. Rev. Lett. {\bf#1},\ #2 (#3)}
\def\RMP #1 #2 #3 {Rev. Mod. Phys. {\bf#1},\ #2 (#3)}
\def\NIM #1 #2 #3 {Nuc. Inst. Meth. {\bf#1},\ #2 (#3)}
\def\ZPC #1 #2 #3 {Z. Phys. {\bf#1},\ #2 (#3)}
\def\EJPC #1 #2 #3 {E. Phys. J. {\bf#1},\ #2 (#3)}
\def\IJMP #1 #2 #3 {Int. J. Mod. Phys. {\bf#1},\ #2 (#3)}
\def\JHEP #1 #2 #3 {J. High En. Phys. {\bf#1},\ #2 (#3)}

\end{document}